\documentclass[11pt]{article}

\usepackage[T1]{fontenc}
\usepackage[utf8]{inputenc}
\usepackage{mathpazo}
\usepackage[scaled=0.92]{helvet}
\usepackage{courier}
\usepackage[margin=1in]{geometry}
\usepackage{amsmath,amssymb,mathtools,bm}
\usepackage{amsthm}
\usepackage{booktabs,longtable,tabularx,array,multirow,threeparttable}
\usepackage{graphicx}
\usepackage{xcolor}
\usepackage{microtype}
\usepackage{enumitem}
\usepackage{caption}
\usepackage{subcaption}
\usepackage{float}
\usepackage{placeins}
\usepackage{ragged2e}
\usepackage{listings}
\usepackage[numbers,sort&compress]{natbib}
\usepackage{xurl}
\usepackage{hyperref}
\usepackage[nameinlink,noabbrev]{cleveref}
\usepackage{lineno}
\usepackage{tikz}
\usetikzlibrary{arrows.meta,positioning,fit,calc,shapes.geometric}

\definecolor{navy}{HTML}{17365D}
\definecolor{teal}{HTML}{0E6B68}
\definecolor{burgundy}{HTML}{8A2432}
\definecolor{paleNavy}{HTML}{EAF0F7}
\definecolor{paleTeal}{HTML}{E8F4F2}
\definecolor{paleRed}{HTML}{F9ECEE}
\definecolor{codeGray}{HTML}{F4F5F7}

\hypersetup{
  colorlinks=true,
  linkcolor=navy,
  citecolor=navy,
  urlcolor=teal,
  pdftitle={DiffSequence-Q: Diffusion Design with Quantum Assistance for Biologically Active Proteins},
  pdfauthor={Ke-Xin Tang, Zhen-Qiang Weng, Kai Yan, Yao-yao Lu, Yu-Ze Zhu, Hang Song, Junjie-He, Jie-Cheng Yang, Xiang Zhao, Lin Yang, Xian-Min Jin}
}

\renewcommand{\arraystretch}{1.15}
\setlist{leftmargin=2em,itemsep=2pt,topsep=3pt}

\newcolumntype{Y}{>{\raggedright\arraybackslash}X}
\newcolumntype{C}{>{\Centering\arraybackslash}X}
\newcolumntype{P}[1]{>{\raggedright\arraybackslash}p{#1}}
\newcommand{\thead}[1]{\textbf{#1}}
\newcommand{\papertablestyle}{\fontsize{9}{11}\selectfont
  \setlength{\tabcolsep}{4pt}\renewcommand{\arraystretch}{1.12}\setlength{\heavyrulewidth}{0.8pt}\setlength{\lightrulewidth}{0.5pt}\setlength{\cmidrulewidth}{0.4pt}}
\newcommand{\papertablenotes}{\fontsize{8}{10}\selectfont}

\newcommand{\R}{\mathbb{R}}
\newcommand{\E}{\mathbb{E}}
\newcommand{\KL}{D_{\mathrm{KL}}}

\newcommand{\softmax}{\operatorname{softmax}}
\newcommand{\Cat}{\operatorname{Cat}}
\newcommand{\Pool}{\operatorname{Pool}}
\newcommand{\diag}{\operatorname{diag}}
\newcommand{\codepath}[1]{\nolinkurl{#1}}

\lstdefinestyle{paperCode}{
  basicstyle=\ttfamily\footnotesize,
  numbers=left,
  numberstyle=\tiny\color{gray},
  frame=single,
  breaklines=true,
  showstringspaces=false,
  backgroundcolor=\color{codeGray},
  columns=fullflexible
}

\title{\textbf{DiffSequence-Q: Diffusion Design with Quantum Assistance for Biologically Active Proteins}}

\newcommand{\AuthorBlock}{%
\begin{minipage}{0.95\textwidth}
\centering
\begin{tabular}{@{}c@{\hspace{1.1em}}c@{\hspace{1.1em}}c@{\hspace{1.1em}}c@{}}
Ke-Xin Tang$^{1}$ & Zhen-Qiang Weng$^{1}$ &
Kai Yan$^{1}$ & Yao-Yao Lu$^{1}$
\end{tabular}\par\vspace{0.35em}
\begin{tabular}{@{}c@{\hspace{1.1em}}c@{\hspace{1.1em}}c@{\hspace{1.1em}}c@{}}
Yu-Ze Zhu$^{1}$ & Hang Song$^{2}$ &
Jun-Jie He$^{1}$ & Jie-Cheng Yang$^{1}$
\end{tabular}\par\vspace{0.35em}
\begin{tabular}{@{}c@{\hspace{1.5em}}c@{\hspace{1.5em}}c@{}}
Xiang Zhao$^{1,\dagger}$ & Lin Yang$^{1,\dagger}$ &
Xian-Min Jin$^{1,2,3,4,5,\dagger}$
\end{tabular}\par\vspace{0.8em}
{\small
$^{1}$TuringQ Co., Ltd., Shanghai 200240, China\par\vspace{0.2em}
$^{2}$Center for Integrated Quantum Information Technologies (IQIT),\par
School of Physics and Astronomy and State Key Laboratory of\par
Photonics and Communications,\par
Shanghai Jiao Tong University, Shanghai 200240, China\par\vspace{0.2em}
$^{3}$Chip Hub for Integrated Photonics Xplore (CHIPX),\par
Shanghai Jiao Tong University, Wuxi 214000, China\par\vspace{0.2em}
$^{4}$Atomology Co., Ltd., Shanghai 200240, China\par\vspace{0.2em}
$^{5}$Hefei National Laboratory, Hefei 230088, China\par
}
\end{minipage}%
}
\author{\AuthorBlock}
\date{}

\begin{document}
\maketitle

\begin{abstract}
\setlength{\parindent}{0pt}
\noindent
Designing protein sequences that satisfy both secondary structure requirements and solubility targets remains challenging. We present DiffSequence-Q, a hybrid multinomial diffusion model conditioned on secondary structure at each residue and a continuous solubility target. To introduce quantum auxiliary supervision during training, we employ a quantum neural network (QNN) or quantum Kolmogorov--Arnold network (QKAN) as an auxiliary solubility prediction head over shared Transformer representations. The auxiliary prediction loss is jointly optimized with the diffusion objective, allowing quantum auxiliary supervision to influence the learned denoising model. Once trained, protein sequences are generated entirely through classical reverse diffusion. On the CATH benchmark, DiffSequence-Q achieved the highest secondary structure recovery among the four evaluated models for complete sequence generation. Evaluation with a separately trained solubility prediction model showed that the best QNN and QKAN configurations achieved mean predicted solubility scores 14.0\% and 22.7\% higher than the multilayer perceptron (MLP) baseline, respectively. Experimental characterization further demonstrated biological activity across three distinct protein families. Selected defensins exhibited pronounced inhibitory activity against \textit{Aspergillus niger}, while xylanases and short-chain dehydrogenases exhibited substantial enzymatic activity. Together, these results support DiffSequence-Q as a framework for integrating structural requirements and solubility targets in protein generation, with experimentally demonstrated biological activity.
\end{abstract}

\begin{center}
\footnotesize
$^{\dagger}$Corresponding authors:
\texttt{zhaoxiang@turingq.com},
\texttt{yanglin@turingq.com},
\texttt{xianmin.jin@sjtu.edu.cn}.
\end{center}

\clearpage

\section{Introduction}

\begin{figure}
    \centering
    \includegraphics[width=\linewidth]{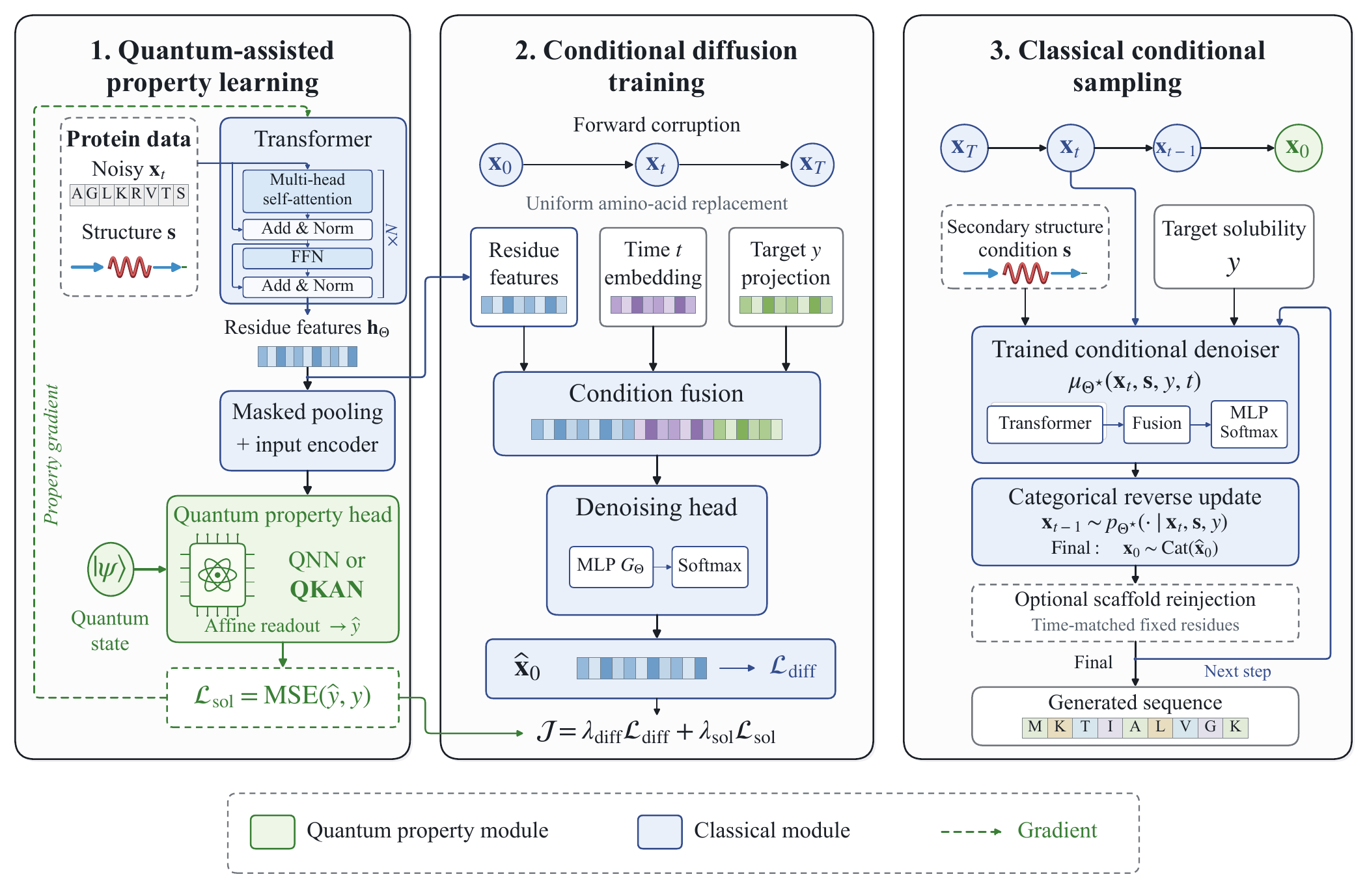}
    \caption{Overview of DiffSequence-Q. The two branches share a Transformer encoder during joint fine-tuning. (1) A QNN or QKAN predicts the original protein's solubility from pooled residue features. (2) The denoising head uses the features at each position, the diffusion step, and the solubility target to predict amino acid probabilities. (3) The trained classical denoiser generates sequences conditioned on the target secondary structure and solubility. For constrained redesign, the sampler preserves selected reference residues in the final sequence.}
    \label{fig:method}
\end{figure}

{Advances in protein structure prediction, exemplified by AlphaFold, have substantially improved our ability to infer the three dimensional structures of proteins from amino acid sequences \citep{Jumper2021AlphaFold}. However, predicting the structure of a given sequence differs from designing a sequence that satisfies specified structural requirements and exhibits desirable biochemical properties \citep{Chu2024ProteinDesign}. The ability to design such proteins is important for applications in biotechnology and medicine. This task remains difficult because the number of possible amino acid sequences grows exponentially with protein length, while structural and biochemical constraints act at different scales. Secondary structure can be specified at individual residues, whereas solubility is a property of the whole sequence. Moreover, satisfying local structural constraints does not necessarily ensure favorable solubility or biological activity. Conditional generative models provide a means of integrating these local and global constraints during sequence design.}

Diffusion models provide one approach to conditional generation through iterative denoising. Denoising diffusion probabilistic models (DDPMs) combine a fixed corruption process with learned reverse transitions for continuous data\citep{Ho2020DDPM}. For categorical data, Hoogeboom et al. formulated multinomial diffusion using uniform replacement noise, an analytic posterior, and a neural estimate of the clean distribution\citep{Hoogeboom2021Multinomial}. {This formulation is particularly suitable for protein sequence generation because amino acids are naturally represented as discrete categories.}

Protein diffusion conditioned on secondary structure has been studied using continuous diffusion models with attention. Ni et al. generated sequences from either global secondary structure composition or secondary structure targets for each residue and found conditioning at each residue to be more accurate and diverse\citep{Ni2023SecondaryDiffusion}. Their Gaussian denoising formulation establishes secondary structure as a condition for inverse design. Other protein generators operate autoregressively, condition on atomic backbones, or generate structures in three dimensions\citep{Madani2023ProGen,Dauparas2022ProteinMPNN,Watson2023RFdiffusion,Ingraham2023Chroma}. {These studies demonstrate progress in controlling structural aspects of protein generation. However, structural conditioning alone does not ensure that generated sequences meet biochemical requirements such as solubility. Incorporating continuous property targets while preserving structural constraints at individual residues therefore remains an important challenge.}

{To address this challenge, }DiffSequence-Q combines conditioning on secondary structure at each residue with a continuous solubility target in a categorical diffusion model. Motivated by hybrid learning architectures that combine classical networks and quantum modules \citep{Mari2020HybridTransfer}, we investigate a differentiable solubility head implemented as either a QNN or a QKAN. The head predicts solubility from a shared Transformer representation, and its loss updates both the property head and the shared encoder. {The quantum property loss therefore influences the learned denoising parameters during joint training, while sequence generation uses the trained classical reverse process, as illustrated in Figure~\ref{fig:method}. We evaluate secondary structure recovery on CATH, compare predicted solubility across different model configurations, and experimentally assess the biological activity of selected protein candidates.}

{Our main contributions are as follows:}

\begin{enumerate}

\item \textbf{Joint conditioning on secondary structure and solubility:}
We propose DiffSequence-Q, a multinomial diffusion framework
that integrates secondary structure constraints at individual
residues with a continuous solubility target through joint
training with quantum property predictors.
On the CATH benchmark, the pretrained diffusion model achieves the highest secondary
structure recovery among the four evaluated methods for complete
sequence generation, outperforming the strongest baseline by
8.96 percentage points in DSSP9 recovery and 4.77 percentage
points in Q3 recovery.

\item \textbf{Scaling Trends with quantum auxiliary heads:}
We investigate how the configurations of quantum auxiliary heads influence the predicted solubility of generated proteins. Increasing the QNN optical mode count and QKAN Fourier order improves mean predicted solubility over selected ranges, with the best configurations outperforming the classical MLP baseline by 14.0\% and 22.7\%, respectively. The observed trends are nonmonotonic, highlighting the importance of auxiliary head configuration.

\item \textbf{Experimental validation of biological activity:}
We validate the biological activity of generated proteins
across three distinct families.
Selected defensins inhibit \textit{Aspergillus niger},
four xylanases exhibit specific activities of
1.92 to 2.21~U/mg and retain 100\% activity after incubation
at 55~$^{\circ}$C for 2~h,
and two short chain dehydrogenases exhibit specific activities
of $55.25 \pm 3.53$ and $53.10 \pm 1.41$~U/mg.

\end{enumerate}

\section{Related Work}

\subsection{Continuous and categorical diffusion}

Diffusion models learn a data distribution by reversing a prescribed corruption process. DDPMs formulate this process for continuous variables, and improved noise schedules control how information is removed across diffusion time steps\citep{Ho2020DDPM,Nichol2021ImprovedDDPM}. For categorical variables, multinomial diffusion uses uniform replacement noise and an analytic posterior over discrete states\citep{Hoogeboom2021Multinomial}. D3PMs extend this construction to structured transition matrices, including transitions based on proximity in embedding space and transitions to absorbing states\citep{Austin2021D3PM}. These choices determine how categories are corrupted and what information the denoiser reconstructs.

Conditional diffusion introduces target information through the denoiser or the reverse sampling procedure. Classifier guidance uses gradients from a classifier to steer reverse diffusion toward a desired class\citep{Dhariwal2021GuidedDiffusion}. Classifier free guidance combines conditional and unconditional denoiser predictions learned with condition dropout\citep{Ho2022CFG}. For inpainting, RePaint reintroduces known regions and alternates forward and reverse transitions to improve consistency with the supplied context\citep{Lugmayr2022RePaint}. These methods provide distinct mechanisms for incorporating target properties and preserving specified parts of a sample.

\subsection{Protein generation with structural conditioning}

Protein generators model sequences, structures, or their joint distribution. ProGen generates sequences using a language model with control tags that specify protein properties, and its generated enzymes were evaluated experimentally\citep{Madani2023ProGen}. ProteinMPNN designs amino acid sequences conditioned on backbone geometry\citep{Dauparas2022ProteinMPNN}. RFdiffusion generates protein backbones under specifications such as functional motifs and binding targets\citep{Watson2023RFdiffusion}. Chroma combines diffusion over protein structures with sequence design and composable constraints on geometry, symmetry, and other protein attributes\citep{Ingraham2023Chroma}.

Secondary structure offers a direct way to specify local structural organization. Ni et al. developed attention based diffusion models conditioned on either overall secondary structure composition or labels aligned to individual residues\citep{Ni2023SecondaryDiffusion}. CPDiffusion-SS represents secondary structure as a graph and generates sequences through diffusion in a latent representation\citep{Hu2024CPDiffusionSS}. These approaches connect sequence design to structural conditions at the level of residues or secondary structure segments.

Discrete diffusion also supports generation directly in sequence space. EvoDiff learns from evolutionary sequence data and supports sequence completion and motif scaffolding\citep{Alamdari2023EvoDiff}. DPLM combines discrete diffusion pretraining with sequence generation and representation learning, and supports conditioning on partial sequences, structure, and classifier predictions\citep{Wang2024DPLM}. DPLM-2 extends this approach by tokenizing backbone geometry and jointly modeling amino acid and structure tokens, enabling both structure prediction and sequence generation conditioned on structure\citep{Wang2025DPLM2}. Together, these models establish sequence diffusion and structural conditioning as complementary components of protein design.

\subsection{Solubility prediction and property control}

Solubility predictors connect amino acid sequences to soluble expression outcomes. DeepSol learns sequence features with convolutional networks and combines them with additional sequence and structural descriptors\citep{Khurana2018DeepSol}. SoluProt uses gradient boosting over sequence characteristics to predict soluble expression in \textit{Escherichia coli} (\textit{E. coli})\citep{Hon2021SoluProt}. NetSolP uses protein language models and evaluates generalization with sequence identity based data partitioning\citep{Thumuluri2022NetSolP}. PLM\_Sol benchmarks several pretrained protein language models and combines ProtT5 embeddings with recurrent and convolutional layers for solubility classification\citep{Zhang2024PLMSol}. Together, these studies support using learned sequence representations for solubility prediction while accounting for expression host and sequence homology during evaluation.

Property predictions can also guide the distribution of generated sequences. Conditioning by adaptive sampling (CbAS) adapts a generative model toward specified properties using probabilistic oracle predictions\citep{Brookes2019CbAS}. DPLM uses classifier guidance to steer sequence generation toward target attributes, including secondary structure\citep{Wang2024DPLM}. Reinforcement learning from experimental feedback (RLXF) adapts protein language models using reward predictors trained on sequence and function measurements, combining supervised adaptation with reinforcement learning\citep{Blalock2026RLXF}. These methods integrate property information through adaptive sampling, guidance during generation, or updates to the generative model.

\subsection{Quantum machine learning and hybrid generative models}

Variational quantum algorithms pair parameterized quantum circuits with classical optimizers, evaluating cost functions from circuit measurements in a hybrid loop across applications including quantum machine learning\citep{Cerezo2021VQA}. Building on parameterized quantum models, Beer et al. train layered quantum channels to learn transformations between quantum states\citep{Beer2020DeepQNN}, and Mari et al. combine pretrained classical feature extractors with trainable quantum circuits and classical readout in a hybrid transfer learning framework\citep{Mari2020HybridTransfer}.

Quantum circuits have also been integrated into attention-based and generative architectures. Cherrat et al. construct quantum vision Transformers using quantum attention and trainable orthogonal layers\citep{Cherrat2024QViT}. Smaldone et al. combine quantum attention scores with classical value matrices and decoder layers to generate SMILES strings conditioned on molecular properties\citep{Smaldone2025HybridTransformer}. Chen et al.'s QD3PM uses quantum channels and trainable denoising circuits to learn joint distributions over discrete data\citep{Chen2026QD3PM}.

End-to-end training of hybrid architectures requires exact gradients of quantum expectation values with respect to both trainable gate parameters and encoded inputs. The parameter-shift rule computes these derivatives from shifted circuit evaluations for gates whose generators have two distinct eigenvalues\citep{Schuld2019ParameterShift}, and generalized parameter-shift rules extend this approach to gates producing expectation values with multiple Fourier frequencies\citep{Wierichs2022GeneralShift}. Combining these circuit derivatives with classical backpropagation via the chain rule enables joint optimization of quantum modules and upstream classical encoders.

This Fourier structure also links parameterized quantum circuits to Kolmogorov--Arnold networks (KANs). Quantum model outputs admit Fourier expansions governed by the data-encoding gates\citep{Schuld2021Encoding}, and in photonic circuits, the input photon number determines the accessible frequencies while the interferometers and readout control the Fourier coefficients\citep{Gan2022FockExpressivity}. Since KANs parameterize network connections with learnable univariate functions\citep{Liu2025KAN}, photonic circuits naturally support Fourier-based QKAN modules alongside multi-mode QNNs, motivating optical mode count and Fourier order as distinct design axes.

Taken together, prior protein diffusion models rarely unify local structural constraints with whole-sequence solubility supervision in discrete sequence space, while hybrid quantum models either serve solely as predictors or incur circuit overhead during iterative sampling. DiffSequence-Q addresses both limitations by pairing a multinomial denoiser conditioned on both structure and solubility with a jointly trained photonic QNN or QKAN head over shared sequence features. Quantum supervision shapes the shared representation during training via interference and Fourier inductive biases, whereas inference relies entirely on classical reverse sampling.

\section{Problem Formulation}

Let $C_{\mathrm{aa}}=20$ be the number of canonical amino acids and $L$ the padded sequence length. The pretraining corpus contains amino acid sequences paired with residue-level DSSP9 secondary structure labels (\cref{sec:datasets}). For training examples with a solubility label, we write
\begin{equation}
 \mathcal D_{\mathrm{sol}}
 =\left\{(\mathbf x_0^{(i)},\mathbf s^{(i)},y^{(i)},\mathbf m^{(i)})\right\}_{i=1}^{N_{\mathrm{sol}}},
\end{equation}
where $N_{\mathrm{sol}}$ is the number of labeled examples. The vector $\mathbf x_0$ represents the amino acid sequence before any diffusion noise is added. At each valid position $\ell$, its entry $x_{0,\ell}\in\{1,\ldots,C_{\mathrm{aa}}\}$ identifies the amino acid at that position. The aligned label $s_\ell$ belongs to the secondary structure alphabet $\mathcal S$, and $y\in[0,1]$ is the sequence-level solubility target. The mask $\mathbf m\in\{0,1\}^{L}$ distinguishes valid residues ($m_\ell=1$) from padding ($m_\ell=0$).

For a design task, the user specifies a target secondary structure sequence $\mathbf s^\star$ and a target solubility $y^\star$. A star on either condition denotes its target value at generation time. The validity mask $\mathbf m$ identifies the residue positions in the padded input. For constrained redesign, a second mask $\mathbf k\in\{0,1\}^L$, with $k_\ell\le m_\ell$, selects the residues to be held at their reference identities $\mathbf x_{\mathrm{known}}$. Let $\Theta$ denote the trainable parameters of the sequence denoiser and $\Theta^\star$ their values after training. The design objective is to sample a complete amino acid sequence from the conditional distribution
\begin{equation}
 p_{\Theta^\star}(\mathbf x_0\mid\mathbf s^\star,y^\star,\mathbf m,\mathbf x_{\mathrm{known}},\mathbf k).
 \label{eq:design-target}
\end{equation}
For full-sequence generation, $\mathbf k=\mathbf 0$ and all valid residues are sampled. For constrained redesign, the sampler enforces the reference residues through \cref{eq:scaffold-redesign}. The validity mask $\mathbf m$ is implicit in the arguments of the denoiser and in the transition probabilities below.

\section{Methods}
\label{sec:methods}
\subsection{Forward multinomial diffusion}

At diffusion step $t$, $\mathbf x_t$ denotes the full sequence of amino acid indices. For a fixed valid position $\ell$, we represent $x_{t,\ell}$ by a column vector $\mathbf v_{t,\ell}$ with $C_{\mathrm{aa}}$ entries. The entry at index $x_{t,\ell}$ is one and all others are zero. Writing one such vector as a row for each valid residue gives a sequence matrix of shape $L\times C_{\mathrm{aa}}$ after padding. Forward corruption and reverse sampling are applied only at residue positions with $m_\ell=1$. Padding positions ($m_\ell=0$) are skipped.

Let $\mathbf u_{\mathrm{aa}}=\mathbf 1/C_{\mathrm{aa}}$ be the uniform probability vector over amino acids. The forward process acts independently at each valid position. At each step, the amino acid index follows the categorical transition of multinomial diffusion~\citep{Hoogeboom2021Multinomial},
\begin{equation}
 q(x_{t,\ell}\mid x_{t-1,\ell})
 =\Cat\!\left(x_{t,\ell}\,\middle|\,
 \alpha_t\mathbf v_{t-1,\ell}+(1-\alpha_t)\mathbf u_{\mathrm{aa}}\right),
 \label{eq:one-step-forward}
\end{equation}
where $\alpha_t=1-\beta_t$ is the probability of copying the previous amino acid. With probability $\beta_t$, an amino acid is drawn uniformly from all $C_{\mathrm{aa}}$ categories, including the previous one. The probability that the amino acid remains unchanged is therefore $\alpha_t+\beta_t/C_{\mathrm{aa}}$. After $t$ steps the transition is
\begin{equation}
 q(x_{t,\ell}\mid x_{0,\ell})
 =\Cat\!\left(x_{t,\ell}\,\middle|\,
 \bar\alpha_t\mathbf v_{0,\ell}+(1-\bar\alpha_t)\mathbf u_{\mathrm{aa}}\right),
 \qquad
 \bar\alpha_t=\prod_{\tau=1}^{t}\alpha_\tau .
 \label{eq:cumulative-forward}
\end{equation}
At $t=0$, no corruption has occurred, so $\bar\alpha_0=1$.

For $T$ diffusion steps, we use a cosine schedule~\citep{Nichol2021ImprovedDDPM} with offset $\delta=0.008$.
\begin{align}
 f(t)&=\cos^2\!\left(\frac{\pi}{2}\frac{t/T+\delta}{1+\delta}\right),\\
 \beta_t&=\min\!\left(1-\frac{f(t)}{f(t-1)},\,0.999\right),
 \qquad
 \bar\alpha_t=\prod_{\tau=1}^{t}(1-\beta_\tau).
 \label{eq:cosine-schedule}
\end{align}

\subsection{Reverse transitions and denoising loss}

Consider a valid position $\ell$ whose clean amino acid $x_{0,\ell}$ and noisy amino acid $x_{t,\ell}$ are known. By Bayes' rule, the posterior probability vector of each candidate amino acid at step $t-1$ is
\begin{equation}
 \widetilde{\bm\theta}_t(\mathbf v_{t,\ell},\mathbf v_{0,\ell})
 =
 \bigl[\alpha_t\mathbf v_{t,\ell}+(1-\alpha_t)\mathbf u_{\mathrm{aa}}\bigr]
 \odot
 \bigl[\bar\alpha_{t-1}\mathbf v_{0,\ell}+(1-\bar\alpha_{t-1})\mathbf u_{\mathrm{aa}}\bigr],
 \label{eq:posterior-unnormalized}
\end{equation}
where $\odot$ denotes elementwise multiplication. Normalizing this vector over the amino acid categories gives an analytic expression for the posterior distribution of $x_{t-1,\ell}$ given $x_{t,\ell}$ and $x_{0,\ell}$.
\begin{equation}
 \begin{aligned}
 \bm\theta_t^{\mathrm{post}}(\mathbf v_{t,\ell},\mathbf v_{0,\ell})
 &=\frac{\widetilde{\bm\theta}_t(\mathbf v_{t,\ell},\mathbf v_{0,\ell})}
 {\mathbf 1^\top\widetilde{\bm\theta}_t(\mathbf v_{t,\ell},\mathbf v_{0,\ell})},\\
 q(x_{t-1,\ell}\mid x_{t,\ell},x_{0,\ell})
 &=\Cat\!\left(x_{t-1,\ell}\,\middle|\,
 \bm\theta_t^{\mathrm{post}}(\mathbf v_{t,\ell},\mathbf v_{0,\ell})\right).
 \end{aligned}
 \label{eq:posterior-normalized}
\end{equation}

The forward process in \cref{eq:one-step-forward} corrupts only amino
acid identities. Once $x_{0,\ell}$ and $x_{t,\ell}$ are given, the
posterior in \cref{eq:posterior-normalized} is determined by the noise
schedule. During generation, $x_{0,\ell}$ is unknown at a position
being designed. To define the learned reverse transition,
let $\mathbf c$ denote the conditions supplied to the denoiser. During
pretraining, the denoiser takes secondary structure as its only condition,
so $\mathbf c=\mathbf s$. When training continues on proteins with
solubility labels, the denoiser also receives the solubility value,
so $\mathbf c=(\mathbf s,y)$. At design time, the user supplies the
target values $\mathbf c^\star=(\mathbf s^\star,y^\star)$.

At step $t$, the neural denoiser reads the full noisy sequence
$\mathbf x_t$ and the conditions $\mathbf c$. It predicts a matrix
$\widehat{\mathbf X}_0\in[0,1]^{L\times C_{\mathrm{aa}}}$ whose valid
rows each sum to one. Each row gives the probabilities of the 20
possible original amino acids at the corresponding position. We denote
row $\ell$, written as a column vector, by
\[
 \widehat{\boldsymbol\pi}_{0,\ell}
 =\bigl(\widehat{\mathbf X}_0[\ell,:]\bigr)^\top .
\]
\Cref{sec:conditional-denoiser} describes how the denoiser uses both
conditions. The aligned structure labels $\mathbf s$ enter the
Transformer encoder. The denoising MLP
combines the resulting residue features with embeddings of the solubility
condition $y$ and the diffusion step $t$. A softmax at each position gives
$\widehat{\mathbf X}_0=\mu_\Theta(\mathbf x_t,\mathbf s,y,t)$.
For either stage and $t\ge2$, replacing the vector for the original
amino acid $\mathbf v_{0,\ell}$ in \cref{eq:posterior-normalized}
with the predicted vector $\widehat{\boldsymbol\pi}_{0,\ell}$ defines
the model's reverse transition,
\begin{equation}
 p_\Theta(x_{t-1,\ell}\mid\mathbf x_t,\mathbf c)
 =\Cat\!\left(x_{t-1,\ell}\,\middle|\,
 \bm\theta_t^{\mathrm{post}}
 (\mathbf v_{t,\ell},\widehat{\boldsymbol\pi}_{0,\ell})\right).
 \label{eq:reverse-kernel}
\end{equation}
At $t=1$, the model draws $x_{0,\ell}$ directly from
$\Cat(\widehat{\boldsymbol\pi}_{0,\ell})$ at each valid position.

Each residue in $\mathbf x_t$ can attend to other residues in the
Transformer, so an amino acid at one position can influence the
predictions at other positions. Given $\mathbf x_t$
and $\mathbf c$, the sampler draws independently across valid positions
within a reverse step. The entire updated sequence provides the input
to the next step.

Both stages use the multinomial diffusion loss of
Hoogeboom et al.~\citep{Hoogeboom2021Multinomial}.
At $t=1$, the reconstruction loss measures the probability assigned to
the original sequence. At $t\ge2$, the denoising loss compares the exact
posterior from \cref{eq:posterior-normalized} with the learned reverse
transition from \cref{eq:reverse-kernel}.
\begin{align}
 \mathcal L_{\mathrm{rec}}
 &=-\log p_\Theta(\mathbf x_0\mid\mathbf x_1,\mathbf c),\\
 \mathcal L_t
 &=\KL\!\left[
 q(\mathbf x_{t-1}\mid\mathbf x_t,\mathbf x_0)
 \;\middle\|\;
 p_\Theta(\mathbf x_{t-1}\mid\mathbf x_t,\mathbf c)
 \right],\qquad 2\le t\le T .
 \label{eq:multinomial-objective}
\end{align}
Both losses are summed over valid residues.\footnote{The entry
$x_{t-1,\ell}$ of $\mathbf x_{t-1}=(x_{t-1,1},\ldots,x_{t-1,L})$
is the amino acid index at valid position $\ell$. The transitions at individual
residues in \cref{eq:reverse-kernel} give
\[
 p_\Theta(\mathbf x_{t-1}\mid\mathbf x_t,\mathbf c)
 =\prod_{\ell:\,m_\ell=1}
 p_\Theta(x_{t-1,\ell}\mid\mathbf x_t,\mathbf c).
\]
The posterior given $\mathbf x_t$ and $\mathbf x_0$ is likewise a product
over valid positions, with one posterior from \cref{eq:posterior-normalized}
at each position.}
Averaging over diffusion steps gives the training objective
\begin{equation}
 \mathcal L_{\mathrm{diff}}
 =\frac{1}{T}\E_q\!\left[
 \mathcal L_{\mathrm{rec}}+\sum_{t=2}^{T}\mathcal L_t
 \right].
 \label{eq:diffusion-loss}
\end{equation}
The expectation is over forward corruption. When generating an entire sequence, these terms follow from
the variational lower bound on the logarithm of its conditional probability. The terminal prior term is constant with respect to
$\Theta$ and is omitted.

\subsection{Denoiser conditioned on secondary structure and solubility}
\label{sec:conditional-denoiser}

Let $E_x$ and $E_s$ denote the learned embeddings of the amino acid
identity and the secondary structure label, respectively. Each maps its
input to a vector in $\R^d$, with $d=128$. The vector
$\operatorname{PE}(\ell)\in\R^d$ is the sinusoidal encoding of residue
position $\ell$. For a given protein $b$, let $t$ denote its diffusion
step. Different proteins in a batch may have different values of $t$.
The input to the first encoder layer at a valid position $\ell$ is
\begin{equation}
 \mathbf h_{\mathrm{in},\ell}^{(b)}
 =E_x(x^{(b)}_{t,\ell})+E_s(s^{(b)}_\ell)+\operatorname{PE}(\ell).
\end{equation}
Let $\Theta=(\Theta_{\mathrm{enc}},\Theta_{\mathrm{den}})$, where
$\Theta_{\mathrm{enc}}$ comprises the shared embeddings and Transformer
encoder parameters, and $\Theta_{\mathrm{den}}$ comprises the parameters
specific to the denoising branch. The encoder attends to valid residues,
as identified by $\mathbf m^{(b)}$. For the sequence of input vectors
$\mathbf h_{\mathrm{in}}^{(b)}$, it gives the residue representations
\begin{equation}
 \mathbf h_{\Theta}^{(b)}
 =\operatorname{Transformer}_{\Theta_{\mathrm{enc}}}
  \!\left(\mathbf h_{\mathrm{in}}^{(b)}\right)
 \in\R^{L\times d}.
 \label{eq:shared-representation}
\end{equation}
The representation $\mathbf h_{\Theta}^{(b)}$ depends on the diffusion step
through the noisy sequence $\mathbf x_t^{(b)}$. When the model uses a solubility target, the denoising head
applies the same multilayer perceptron (MLP) at every residue. For
sample $b$, it combines each residue representation with the embedding
$\mathbf e_t$ of the diffusion step and the embedding
$\mathbf e_y^{(b)}$ of the solubility target. Both embeddings are shared
across the sequence. This gives the
residue logits and, after a softmax, the probability matrix
\begin{align}
 \bm\ell_{\Theta,\ell}^{(b)}
 &=\operatorname{MLP}_{\Theta_{\mathrm{den}}}\!\left(
 [\mathbf h_{\Theta,\ell}^{(b)};\mathbf e_t;\mathbf e_y^{(b)}]\right)
 \in\R^{C_{\mathrm{aa}}},\\
 \widehat{\mathbf X}_0^{(b)}
 &=\mu_\Theta(\mathbf x_{t}^{(b)},\mathbf s^{(b)},y^{(b)},t)
 =\softmax\!\left(\bm\ell_\Theta^{(b)}\right)
 \in[0,1]^{L\times C_{\mathrm{aa}}}.
 \label{eq:denoiser}
\end{align}
The brackets in the MLP input denote concatenation. Stacking the
residue logits as rows gives $\bm\ell_\Theta^{(b)}$. The softmax acts
across amino acid classes within each row. The symbol $\mu_\Theta$
denotes the full denoiser, from input embeddings to these probabilities. For a batch of $B$ sequences the output has
shape $B\times L\times C_{\mathrm{aa}}$.

\subsection{Auxiliary solubility heads}
\label{sec:solubility-heads}

Each model uses the denoising MLP described in \cref{sec:conditional-denoiser}.
A separate auxiliary head predicts one solubility value for the entire
sequence. It reads the shared Transformer features in
\cref{eq:shared-representation}, before the denoising branch introduces
the solubility condition. Its input is the mean over valid residues,
\begin{equation}
 \mathbf u^{(b)}=\Pool_{\mathbf m^{(b)}}(\mathbf h_{\Theta}^{(b)})
 =\frac{\sum_{\ell=1}^{L}m_{\ell}^{(b)}\mathbf h_{\Theta,\ell}^{(b)}}
 {\max(1,\sum_{\ell=1}^{L}m_{\ell}^{(b)})}\in\R^{128}.
 \label{eq:masked-pooling}
\end{equation}
The head takes this vector as its sole input and predicts a scalar,
\begin{equation}
 \widehat y^{(b)}=\mathcal H_\omega(\mathbf u^{(b)}),
 \qquad \mathcal H_\omega:\R^{128}\to\R.
 \label{eq:solubility-readout}
\end{equation}
The baseline uses an MLP for $\mathcal H_\omega$. In the two quantum
variants, $\omega$ comprises the parameters of the head's input encoder,
its QNN or QKAN module, and its affine readout
(\cref{app:four-mode-unitary,app:photonic-kan-circuit}).
For protein $b$, the input varies with its diffusion step $t$, while the regression
target $y^{(b)}$ is the solubility label of the original protein. This
auxiliary task asks the shared encoder to retain sequence information
that is predictive of solubility after corruption. Related protein
design work trains property predictors on corrupted sequences using
features shared with a denoiser~\citep{Gruver2023GuidedDiscrete}.

\subsection{Pretraining and joint training with solubility labels}
\label{sec:e2e-guidance}

We first pretrained the denoiser on the protein sequences and DSSP9
annotations described in \cref{sec:datasets}. The model learned to
recover the amino acid sequence given the secondary structure at each
residue. We used the diffusion loss in \cref{eq:diffusion-loss} with
$\mathbf c=\mathbf s$.

We then continued training on proteins with solubility labels
(\cref{sec:solubility-data}). We compared three architectures for the
auxiliary solubility head, an MLP, a QNN, and a QKAN.
The original sequence label $y^{(b)}$ serves two roles. It conditions
the denoising head in \cref{eq:denoiser} and supplies the regression
target for the auxiliary prediction $\widehat y^{(b)}$ in
\cref{eq:solubility-readout}. The shared encoder and the auxiliary head
are updated jointly at this stage.

For a labeled batch, let $\widehat{\mathcal L}_{\mathrm{diff}}$ denote
the batch estimate of \cref{eq:diffusion-loss} with
$\mathbf c^{(b)}=(\mathbf s^{(b)},y^{(b)})$ for each example.
For each example $b$, both branches use the same noisy sequence
$\mathbf x_t^{(b)}$, where $t$ is the step sampled for that example.
If $t$ is drawn uniformly from $\{1,\ldots,T\}$, the loss at that step
is an unbiased estimate of the time average in \cref{eq:diffusion-loss}.
It is the reconstruction loss at $t=1$ and the corresponding KL
divergence at $t\ge2$.
We give the auxiliary loss less weight when the noise level is high
because the corrupted sequence retains less information about the
original amino acids. A related weighting scheme has been used for property
prediction from noisy ligands~\citep{Cremer2024PILOT}.
Writing $w^{(b)}\geq0$ for the effective weight at the noise level
of sample $b$, the auxiliary loss and the joint training objective are
\begin{align}
 \widehat{\mathcal L}_{\mathrm{sol}}
 &=\frac{1}{B}\sum_{b=1}^{B}
 w^{(b)}\bigl(\widehat y^{(b)}-y^{(b)}\bigr)^2,
 \label{eq:batch-solubility-mse}\\
 \widehat{\mathcal J}(\Theta,\omega)
 &=\lambda_{\mathrm{diff}}\widehat{\mathcal L}_{\mathrm{diff}}(\Theta)
 +\lambda_{\mathrm{sol}}\widehat{\mathcal L}_{\mathrm{sol}}(\Theta,\omega).
 \label{eq:joint-objective}
\end{align}
The expectation of the batch loss $\widehat{\mathcal J}$ gives the joint
objective $\mathcal J$ in \cref{fig:method}. For a batch of $B$ proteins,
we form a matrix $\mathbf U\in\mathbb R^{B\times128}$ by placing the
pooled features of one protein in each row. Row $b$ is
$(\mathbf u^{(b)})^\top$, where $\mathbf u^{(b)}$ is the mean of the
residue features over the valid positions in protein $b$
(\cref{eq:masked-pooling}). The solubility head predicts one value per
protein, giving the vector
$\widehat{\mathbf y}=(\widehat y^{(1)},\ldots,\widehat y^{(B)})^\top$.
The gradient with respect to the denoiser parameters is
\begin{equation}
 \nabla_\Theta\widehat{\mathcal J}
 =\lambda_{\mathrm{diff}}\nabla_\Theta\widehat{\mathcal L}_{\mathrm{diff}}
 +\lambda_{\mathrm{sol}}\,
 J_{\mathbf U,\Theta}^{\top}
 J_{\widehat{\mathbf y},\mathbf U}^{\top}
 \nabla_{\widehat{\mathbf y}}\widehat{\mathcal L}_{\mathrm{sol}},
 \label{eq:generator-gradient}
\end{equation}
where $J_{\mathbf a,\mathbf b}=\partial\mathbf a/\partial\mathbf b$.
The second term acts on $\Theta_{\mathrm{enc}}$ through the pooled
encoder features. Since $\mathbf U$ is independent of the parameters
specific to the denoising head,
$\nabla_{\Theta_{\mathrm{den}}}\widehat{\mathcal L}_{\mathrm{sol}}=0$.
For sample $b$, the encoder gradient derived in
\cref{eq:photonic-encoder-gradient} enters the joint gradient with weight
$\lambda_{\mathrm{sol}}w^{(b)}/B$. Derivatives through the quantum
modules are given in \cref{app:photonic-parameter-shift}.

\subsection{Reverse generation and constrained redesign}

Sampling starts by drawing each valid amino acid index in $\mathbf x_T$ independently from the uniform distribution. For $t=T,\ldots,2$, the denoiser takes the current sequence $\mathbf x_t$, the target conditions $(\mathbf s^\star,y^\star)$, and the step index $t$ as inputs. It predicts a probability vector for each valid position, and the sampler draws the next amino acid indices using \cref{eq:reverse-kernel}. At $t=1$, it draws each index directly from its predicted clean distribution. One generated sequence is obtained by following one sampled trajectory from $\mathbf x_T$ to $\mathbf x_0$, using only the trained denoiser $\mu_{\Theta^\star}$ at every step.

For constrained redesign, the sampler applies the two reference inputs in \cref{eq:design-target} after each denoising step. The mask $\mathbf k$ selects the fixed positions, and $(\mathbf x_{\mathrm{known}})_\ell$ gives the reference amino acid at each selected position $\ell$. Let $x_{t-1,\ell}^{\mathrm{generated}}$ denote the amino acid drawn by the reverse sampler at position $\ell$. The update is
\begin{equation}
 x_{t-1,\ell}\leftarrow
 \begin{cases}
  x_{t-1,\ell}^{\mathrm{known}}, & k_\ell=1,\\
  x_{t-1,\ell}^{\mathrm{generated}}, & k_\ell=0 .
 \end{cases}
 \label{eq:scaffold-redesign}
\end{equation}
For $k_\ell=1$, the sampler draws $x_{t-1,\ell}^{\mathrm{known}}$ from the forward distribution in \cref{eq:cumulative-forward} at step $t-1$, using $(\mathbf x_{\mathrm{known}})_\ell$ as the initial amino acid. At $t=1$, this gives $x_{0,\ell}=(\mathbf x_{\mathrm{known}})_\ell$ exactly. At earlier updates, the resulting sequence $\mathbf x_{t-1}$ is the input to the next denoising step. Optional alternating forward and reverse steps follow the resampling procedure of RePaint~\citep{Lugmayr2022RePaint}.

\section{Experimental Design}
\label{sec:experiments}

\subsection{Pretraining corpus and CATH benchmark}
\label{sec:datasets}

For pretraining (\cref{sec:e2e-guidance}), we assembled amino acid
sequences from the ProteinMPNN dataset, DSSP Bank, and AlphaFold DB v6.
We applied DSSP to the corresponding structures to assign each valid
residue one of nine secondary structure states (DSSP9).

Sequences of 50--800 residues were retained and
duplicate pairs of sequences and their annotations were removed, yielding 1,523,026 pairs
spanning 922,230 unique amino acid sequences. To reduce potential information
leakage from homologous sequences across data splits, sequences were clustered
using MMseqs2 with a minimum sequence identity of 30\% and a coverage threshold
of 80\%. Entire sequence clusters were assigned to the training, validation,
or test split. Validation data were kept separate from training and used to monitor
training and guide model selection. Statistics of the pretraining corpus are
summarized in Table~\ref{tab:dataset_statistics}.

\begin{table}[htbp]
\centering
\papertablestyle
\begin{threeparttable}
\caption{Statistics of the pretraining corpus. Sequence lengths are reported in amino acid residues.}
\label{tab:dataset_statistics}
\begin{tabular*}{\linewidth}{@{\extracolsep{\fill}}lrrr@{}}
\toprule
\thead{Split} & \thead{Samples} & \thead{Length range} & \thead{Mean length} \\
\midrule
Training & 1,522,826 & 50--800 & 300.12 \\
Validation & 100 & 52--684 & 293.95 \\
Test & 100 & 51--733 & 354.99 \\
\midrule
Total & 1,523,026 & 50--800 & 300.12 \\
\bottomrule
\end{tabular*}
\begin{tablenotes}[flushleft]\papertablenotes
\item[] Each sample is a unique pairing of a sequence and its annotation.
\end{tablenotes}
\end{threeparttable}
\end{table}

We used CATH v4.2 to evaluate the generation of amino acid sequences
given a target secondary structure (\cref{sec:masking-results}). The standard CATH v4.2 test split introduced by
Ingraham et al.~\cite{ingraham2019generative} contains 1,120 protein
chains. \Cref{tab:protein-generation} reports the comparison.

\FloatBarrier

\subsection{Data for joint training with solubility targets}
\label{sec:solubility-data}

Two separate solubility datasets are used for fine-tuning the generative model and for training the external solubility evaluator (\cref{tab:data}). To fine-tune the pretrained DiffSequence-Q model with joint property supervision (\cref{sec:e2e-guidance}), we use the 100--800-residue dataset (99,903 retained of 100,000 records), in which each protein has residue-level DSSP9 tokens (with padding excluded by the validity mask) and a continuous solubility score $y\in[0,1]$ predicted by the pretrained Protein-Sol model. The continuous score both conditions the denoiser and supervises the auxiliary QNN, QKAN, or MLP head. To evaluate generated sequences independently, we train an external ESM++ solubility predictor (\cref{app:solubility-evaluator}) on the cleaned DeepSol \textit{E.~coli} dataset~\citep{Khurana2018DeepSol}, which retains 66,940 unique sequences with experimental binary solubility labels $y\in\{0,1\}$ (60,246 for training and 6,694 for validation). The MLP, QNN, and QKAN variants use the same denoising backbone, with 128-dimensional embeddings, a hidden dimension of 256, 12 Transformer layers, eight attention heads, a maximum sequence length of 800, and 1,000 diffusion steps. All three variants use the same solubility training dataset and are trained using Adam with a learning rate of \(10^{-4}\), a batch size of 64, 30 epochs, and a random seed of 1024. A two-epoch linear learning-rate warmup is applied to all variants. In the transfer-learning implementation, pretrained parameters are loaded when their names and tensor shapes match those of the target model; newly introduced parameters retain their architecture-specific default initialization. Furthermore, the current QNN and QKAN training implementations apply gradient clipping with a maximum norm of 1.0. 

\begin{table}[htbp]
\centering
\papertablestyle
\begin{threeparttable}
\caption{Statistics of the solubility datasets for generative fine-tuning and external evaluator training. Lengths are reported in amino acid residues, and $\bar y$ denotes the mean solubility target.}
\label{tab:data}
\begin{tabular*}{\linewidth}{@{\extracolsep{\fill}}lrrrrr@{}}
\toprule
\thead{Dataset} & \thead{Records} & \thead{Retained} & \thead{Length (median/mean)} & $\bm{\bar y}$ & \thead{Unique seq.}\\
\midrule
Generative fine-tuning (100--800) & 100,000 & 99,903\tnote{a} & 336 / 373.36 & 0.49626 & 96,109\\
Evaluator training (DeepSol \textit{E.~coli}) & 71,421 & 66,940\tnote{b} & 276 / 300.38 & 0.41521 & 66,940\\
\bottomrule
\end{tabular*}
\begin{tablenotes}[flushleft]\papertablenotes
\item[a] Computationally predicted continuous targets $y\in[0,1]$; sequences of 799 and 800 residues were excluded.
\item[b] Cleaned experimental binary labels $y\in\{0,1\}$ from DeepSol\citep{Khurana2018DeepSol}, split into 60,246 training and 6,694 validation sequences. 
\end{tablenotes}
\end{threeparttable}
\end{table}

\FloatBarrier
\subsection{Evaluation Tasks and Metrics}
\label{sec:evaluation-protocol}
\label{sec:masking-evaluation}

Generation was evaluated in terms of secondary structure recovery, structural confidence, and predicted solubility.

\paragraph{Conditional generation and structural recovery.}
Three masking settings were evaluated: the full amino acid sequence, its
first half, and its last half. Unmasked residues and complete
structural conditions were retained in each setting.
Sequences were folded with ESMFold and annotated with DSSP.
DSSP9 recovery measures exact agreement with target labels. Q3 recovery
uses the mapping $\{\texttt{H},\texttt{G},\texttt{I}\}\to\texttt{H}$,
$\{\texttt{E},\texttt{B}\}\to\texttt{E}$, and
$\{\texttt{T},\texttt{S},\texttt{P},\texttt{-}\}\to\texttt{C}$,
corresponding to helix, strand, and coil, respectively.
Both metrics report the percentage of matching positions over the whole
sequence and, for partial masking, the redesigned and fixed regions.
Fixed regions retain amino acid identities,
but their predicted secondary structures may change.

\paragraph{Structural confidence.}
ESMFold mean pLDDT quantifies local structural confidence.
Higher values indicate greater confidence.

\paragraph{Predicted solubility.}
The same external predictor (\cref{app:solubility-evaluator}) scored the
sequences generated by each model. We report the mean, minimum, and
maximum predicted solubility within each group (\cref{sec:trained-model-results}). For the generation experiments reported, each configuration generates 100 sequences under the same fixed conditions. Specifically, \(\mathbf{s}^{\star}\) is the complete 527-residue secondary-structure label sequence supplied for human monoamine oxidase (MAO; UniProt P21397). The target solubility is set to \(y^{\star}=1.0\) for every sample and remains fixed throughout reverse diffusion. Generation performs constrained redesign of 179 residues at positions 6–13, 33, 38, 49, 52–53, 62–63, 87–94, 106–111, 118–134, 139–150, 195–204, 229–235, 291–300, 343–354, 370–377, 383–392, and 464–527, using one-based indexing. The remaining 348 residues retain their reference amino acid identities. All generated sequences are scored using the same external ESM++ based solubility predictor.

\subsection{Characterization of plant defensins}
\label{sec:plant-defensins}

For each candidate plant defensin, we cloned a sequence encoding two tandem copies of the defensin joined by a flexible linker into pET-28a(+) between the \textit{Nco}I and \textit{Xho}I restriction sites. The resulting fusion protein carried a C-terminal \(6\times\)His tag. Competent \textit{E. coli} BL21(DE3) cells were transformed with the recombinant plasmids by heat shock, and positive clones were selected for protein expression. Cultures were grown in \(50\,\mathrm{mL}\) of terrific broth (TB) at \(37\,^{\circ}\mathrm{C}\). When the \(\mathrm{OD}_{600}\) reached 0.6--0.8, expression was induced by adding isopropyl \(\beta\)-D-1-thiogalactopyranoside (IPTG) to a final concentration of \(0.1\,\mathrm{mM}\), followed by incubation at \(20\,^{\circ}\mathrm{C}\) for 24~h. The harvested cells were resuspended in potassium phosphate buffer (pH~7.0) and disrupted by sonication on ice for 10~min. The lysates were centrifuged at 12,000~rpm for 10~min under refrigerated conditions, and the supernatants were collected as crude protein preparations.

Antifungal activity against \textit{Aspergillus niger} was evaluated using a plate-based growth inhibition assay. Protein preparations from cells expressing defensin constructs Defensin A, B and C were applied to agar plates inoculated with \textit{A.~niger}. After incubation at \(28\,^{\circ}\mathrm{C}\) for 24~h, the plates were examined for zones of growth inhibition. All experiments were performed in triplicate.

\subsection{Characterization of xylanase}
\label{sec:xylanase}

The coding sequence of the candidate xylanase was cloned into pET-21a between the \textit{Nde}I and \textit{Xho}I restriction sites to generate a construct encoding a C-terminal \(6\times\)His-tagged protein. Competent \textit{E.~coli} BL21(DE3) cells were transformed with the recombinant plasmid by heat shock, and positive clones were selected for protein expression.

Xylanase activity was evaluated by quantifying the reducing sugars released from xylan using the 3,5-dinitrosalicylic acid (DNS) assay. The reaction mixture contained \(1\,\mathrm{g\,L^{-1}}\) xylan and \(0.1\,\mathrm{mg\,mL^{-1}}\) protein in potassium phosphate buffer (pH~5.8). After incubation at \(55\,^{\circ}\mathrm{C}\) for 30~min, the samples were heated at \(95\,^{\circ}\mathrm{C}\) for 10~min and centrifuged at 12,000~rpm for 2~min at room temperature. Reducing sugars were then quantified using the DNS assay. One unit (U) of xylanase activity was defined as the amount of enzyme that released \(1\,\mu\mathrm{mol}\) of reducing sugars per minute under the specified assay conditions.

To assess thermal stability, the enzyme preparation was preincubated at \(55\,^{\circ}\mathrm{C}\) for 2~h. Residual activity was then measured under the same reaction conditions using the DNS assay. All experiments were performed in triplicate.

\subsection{Characterization of SDR}
\label{sec:sdr}

The coding sequence of the candidate short-chain dehydrogenase/reductase (SDR) was cloned into pRSFDuet-1 between the \textit{BamH}I and \textit{Xho}I restriction sites to generate a construct encoding an N-terminal \(6\times\)His-tagged protein. Competent \textit{E.~coli} BL21(DE3) cells were transformed with the recombinant plasmid by heat shock, and positive clones were selected for protein expression. The recombinant protein was purified by nickel-affinity chromatography using a GenScript column and an \"{A}KTA chromatography system. Protein concentrations were determined using the bicinchoninic acid (BCA) assay.

\(\mathrm{NADP}^{+}\)-dependent catalytic activity was evaluated using ethanol as a model substrate. Assays were conducted in potassium phosphate buffer (pH~7.0) at a final protein concentration of \(0.1\,\mathrm{mg\,mL^{-1}}\). Activity was assessed by measuring the reduction of \(\mathrm{NADP}^{+}\) to \(\mathrm{NADPH}\) during substrate oxidation. One unit (U) of activity was defined as the amount of enzyme that catalyzed the reduction of \(1\,\mu\mathrm{mol}\) of \(\mathrm{NADP}^{+}\) per minute under the specified assay conditions. Specific activity was expressed as \(\mathrm{U\,mg^{-1}}\) protein. All experiments were performed in triplicate.

\section{Results}

\subsection{Secondary Structure Recovery of the Pretrained Diffusion Model}
\label{sec:masking-results}
\label{sec:compared-models}

Results for pretrained models are reported for DiffSequence-Q,
ProteinDiffusionGenerator~\cite{Ni2023SecondaryDiffusion},
DPLM-2~\cite{Wang2025DPLM2}, and CPDiffusion-SS~\cite{Hu2024CPDiffusionSS}.
DiffSequence-Q and ProteinDiffusionGenerator received complete secondary structure labels aligned
to each residue (DSSP9 for DiffSequence-Q). DPLM-2 received
discrete structure tokens encoded from the full target backbone, whereas
CPDiffusion-SS received a secondary structure graph containing segment
types, centroids in three dimensions, and spatial edges.

\begin{table}[!htbp]
\centering
\begingroup
\papertablestyle
\begin{threeparttable}
\caption{Pretrained model evaluations under three amino acid masking settings.}
\label{tab:protein-generation}
\begin{tabular*}{\linewidth}{@{\extracolsep{\fill}}lrrrrrrr@{}}
\toprule
 & \multicolumn{6}{c}{\thead{Secondary structure recovery (\%)}}
 & \multicolumn{1}{c}{\thead{Mean confidence}} \\
\cmidrule(lr){2-7}\cmidrule(lr){8-8}
\thead{Model} & \multicolumn{2}{c}{\thead{Whole sequence}}
 & \multicolumn{2}{c}{\thead{Redesigned region}}
 & \multicolumn{2}{c}{\thead{Fixed region}} & \multicolumn{1}{c}{\thead{pLDDT}} \\
\cmidrule(lr){2-3}\cmidrule(lr){4-5}\cmidrule(lr){6-7}
 & \thead{DSSP9} & \thead{Q3} & \thead{DSSP9} & \thead{Q3} & \thead{DSSP9} & \thead{Q3} & \\
\midrule
\multicolumn{8}{@{}l}{\textit{CATH: masking the whole sequence}} \\
\addlinespace[2pt]
\textbf{DiffSequence-Q (ours)} & \textbf{43.71} & \textbf{60.57} & \textbf{43.71} & \textbf{60.57} & N/A & N/A & 43.13 \\
ProteinDiffusionGenerator & \underline{34.75} & \underline{55.80} & \underline{34.75} & \underline{55.80} & N/A & N/A & 30.28 \\
DPLM-2 & 25.31 & 37.15 & 25.31 & 37.15 & N/A & N/A & \textbf{81.15} \\
CPDiffusion-SS & 31.47 & 42.36 & 31.47 & 42.36 & N/A & N/A & \underline{47.03} \\
\midrule
\multicolumn{8}{@{}l}{\textit{CATH: masking the first half}} \\
\addlinespace[2pt]
\textbf{DiffSequence-Q (ours)} & \textbf{63.78} & \textbf{77.19} & \textbf{49.58} & \textbf{66.61} & \underline{78.06} & \underline{87.83} & 56.68 \\
ProteinDiffusionGenerator & \underline{61.38} & \underline{74.48} & 37.64 & 55.42 & \textbf{85.21} & \textbf{93.61} & \underline{62.70} \\
DPLM-2 & 61.15 & 72.77 & \underline{44.59} & \underline{57.95} & 77.81 & 87.67 & \textbf{74.47} \\
CPDiffusion-SS & 54.50 & 66.08 & 33.03 & 46.46 & 76.10 & 85.81 & 56.58 \\
\midrule
\multicolumn{8}{@{}l}{\textit{CATH: masking the last half}} \\
\addlinespace[2pt]
\textbf{DiffSequence-Q (ours)} & 62.65 & \underline{75.72} & \underline{46.76} & \textbf{63.11} & 78.45 & 88.26 & 54.30 \\
ProteinDiffusionGenerator & \underline{63.00} & \textbf{76.97} & 33.48 & 58.37 & \textbf{92.41} & \textbf{95.51} & \underline{61.37} \\
DPLM-2 & \textbf{63.33} & 74.25 & \textbf{47.68} & \underline{59.95} & \underline{78.90} & \underline{88.48} & \textbf{74.76} \\
CPDiffusion-SS & 53.01 & 63.95 & 29.02 & 41.15 & 76.87 & 86.63 & 56.41 \\
\bottomrule
\end{tabular*}
\begin{tablenotes}[flushleft]\papertablenotes
\item[] N/A: no fixed region. Bold and underlined values indicate the highest and second-highest scores, respectively, among the four models for each column and masking setting.
\end{tablenotes}
\end{threeparttable}
\endgroup
\end{table}

Table~\ref{tab:protein-generation} summarizes the three masking settings.
DiffSequence-Q achieved Q3 recovery rates of 60.57\%, 66.61\%, and 63.11\%
in the redesigned region when masking the whole sequence, its first half,
and its last half, respectively. Among the CATH results, DPLM-2
had the highest mean pLDDT in all three settings.

\FloatBarrier

\subsection{Predicted Solubility with Simulated Quantum Auxiliary Heads}
\label{sec:trained-model-results}

We evaluate how different auxiliary solubility prediction heads affect the predicted solubility of redesigned protein sequences. During joint fine tuning, an MLP, QNN, or QKAN head predicts solubility from the shared Transformer representations, and its prediction loss provides additional supervision to the encoder alongside the diffusion objective. The quantum modules considered in this section are evaluated through numerical simulation. Once training is complete, sequence generation proceeds entirely through classical reverse diffusion.

The three variants share the same diffusion backbone and the core training settings detailed in \cref{sec:solubility-data}. These include a Transformer with 12 layers and eight attention heads, 128 dimensional embeddings, a hidden dimension of 256, and 1,000 diffusion steps. All variants use the same solubility training dataset and Adam optimizer, with a learning rate of $10^{-4}$, a batch size of 64, 30 training epochs, a two epoch linear warmup, and a random seed of 1024. Pretrained parameters are transferred where their names and tensor shapes match; new parameters use the default initialization of the corresponding architecture. The reported QNN and QKAN implementations additionally employ gradient clipping with a maximum norm of 1.0.

\begin{figure}[htbp]
  \centering
  \includegraphics[width=0.94\linewidth]{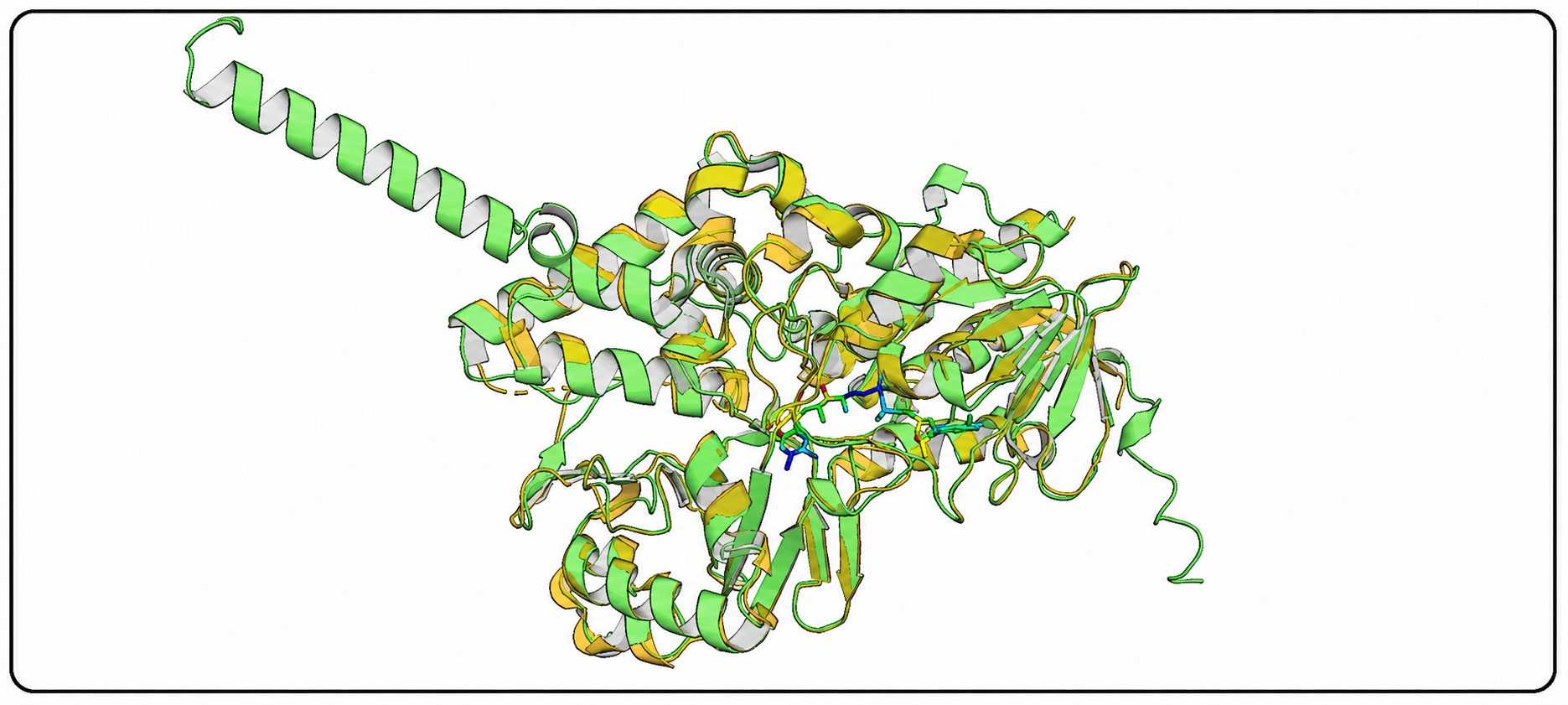}
  \caption{MAO sequence and predicted protein structure.}
  \label{fig:mao-structure}
\end{figure}

Human monoamine oxidase (MAO, UniProt: P21397, 527 residues) serves as the reference protein (Figure~\ref{fig:mao-structure}). For every configuration, the target secondary structure $\mathbf{s}^{\star}$ is the same complete 527 residue secondary structure label sequence, and the target solubility is fixed at $y^{\star}=1.0$. Generation uses constrained redesign at 179 designated positions, while the other 348 residues retain their reference identities, as described in \cref{sec:masking-evaluation}. Each configuration generates 100 candidate sequences under these fixed conditions. All candidates are evaluated using the same independently trained solubility predictor based on ESM++ embeddings (\cref{app:solubility-evaluator}). The same MLP reference is used in the QNN and QKAN comparisons.

\subsubsection{Scaling Trends with QNN Optical Mode Count}
\label{sec:mode-solubility}

\begin{figure}[!htbp]
  \centering
  \includegraphics[width=\linewidth]{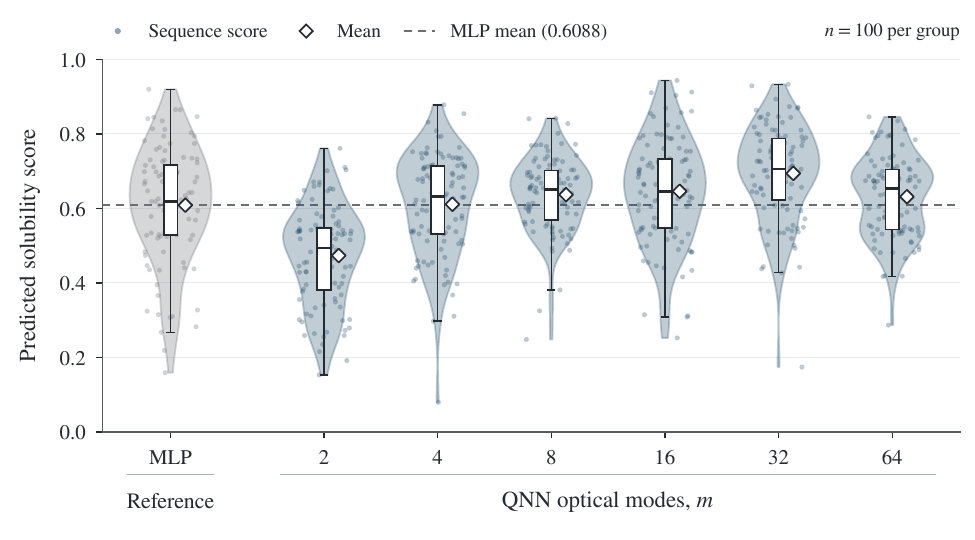}
  \caption{Predicted solubility of QNN heads across different numbers of optical modes. Each group, including the MLP reference, contains 100 generated sequences scored with the same external predictor. The MLP sequences and scores are shared with Figure~\ref{fig:qkan-order-solubility}. Violins show Gaussian kernel density estimates with Scott's bandwidth, restricted to the observed score ranges and scaled to equal maximum widths. Dots show individual sequence scores with horizontal jitter. Boxes span the interquartile range (IQR), central bars mark medians, and whiskers extend to the most extreme observations within 1.5 IQR of the box. Diamonds mark means. The dashed horizontal line denotes the MLP mean (0.6088).}
  \label{fig:mode-solubility}
\end{figure}

We first examine how optical mode count relates to predicted solubility across six QNN configurations with 2, 4, 8, 16, 32, and 64 modes. \Cref{tab:solubility,fig:mode-solubility} summarize the scores of the 100 redesigned sequences per configuration and compare them with the classical MLP reference.

\begin{table}[!htbp]
  \centering
  \begingroup
  \papertablestyle
  \begin{threeparttable}
  \caption{Predicted solubility scores and parameter counts for six QNN configurations and the MLP reference, with 100 generated sequences per group.}
  \label{tab:solubility}
  \begin{tabular*}{\linewidth}{@{\extracolsep{\fill}}lcrrrrr@{}}
    \toprule
    & & \multicolumn{3}{c}{\thead{Predicted score}}
    & \multicolumn{2}{c}{\thead{Trainable parameters}} \\
    \cmidrule(lr){3-5}\cmidrule(lr){6-7}
    \thead{Head} & \thead{Modes} & \thead{Mean} & \thead{Min.} & \thead{Max.} & \thead{Total} & \thead{Circuit} \\
    \midrule
    MLP & --- & 0.6088 & 0.1593 & 0.9197 & 18,649 & --- \\
    \midrule
    \multirow{6}{*}{QNN} & 2  & 0.4736 & 0.1532 & 0.7612 & 16,652 & 8 \\
        & 4  & 0.6112 & 0.0796 & 0.8780 & 16,936 & 32 \\
        & 8  & 0.6369 & 0.2483 & 0.8420 & 17,552 & 128 \\
        & 16 & \underline{0.6461} & \underline{0.2525} & \textbf{0.9435} & 18,976 & 512 \\
        & 32 & \textbf{0.6941} & 0.1742 & \underline{0.9330} & 22,592 & 2,048 \\
        & 64 & 0.6315 & \textbf{0.2870} & 0.8457 & 32,896 & 8,192 \\
    \bottomrule
  \end{tabular*}
    \begin{tablenotes}[flushleft]
    \papertablenotes
    \item[] Bold and underlined values denote the highest and second highest QNN entries in each score column.
    Total counts refer to the auxiliary prediction head and include circuit parameters.
    The photonic QNN architecture is detailed for four optical modes in \cref{app:four-mode-unitary}.
    \end{tablenotes}
  \end{threeparttable}
  \endgroup
\end{table}

The mean predicted solubility increases from 0.4736 with 2 modes to 0.6112, 0.6369, and 0.6461 with 4, 8, and 16 modes, respectively. The highest mean score of 0.6941 occurs with 32 modes, corresponding to a 14.0\% increase relative to the MLP reference of 0.6088. With 64 modes, however, the mean decreases to 0.6315, even as the number of trainable parameters in the auxiliary head increases from 22,592 to 32,896.

The distributions in \cref{fig:mode-solubility} show that the configuration with the highest mean is not the best on every reported statistic. The highest observed maximum score, 0.9435, occurs with 16 modes, while the highest observed minimum score, 0.2870, occurs with 64 modes. The 32 mode configuration has an observed minimum of 0.1742 and maximum of 0.9330, illustrating the variation among its 100 redesigned sequences.

At a smaller scale, the 4 mode QNN achieves a mean score of 0.6112, close to the MLP reference, with 16,936 trainable head parameters rather than 18,649. This is a 9.19\% reduction in the auxiliary head parameter count. Overall, the QNN results exhibit an increasing trend in mean predicted solubility from 2 to 32 modes, followed by a decline at 64 modes. The trend is therefore not monotonic, and a larger optical mode count does not consistently yield higher predicted solubility.

\subsubsection{Scaling Trends with QKAN Fourier Order}
\label{sec:qkan-results}
\label{sec:qkan-generation}

\begin{figure}[!htbp]
  \centering
  \includegraphics[width=\linewidth]{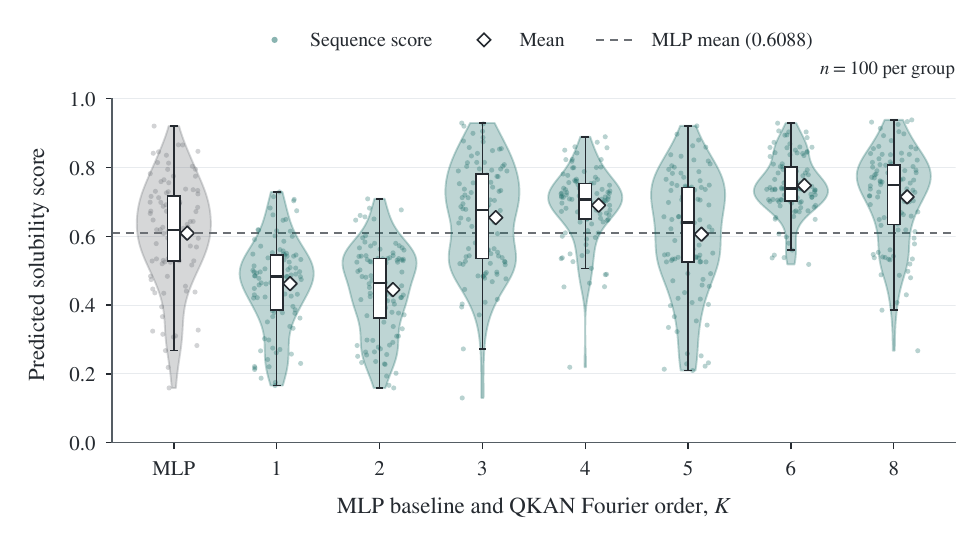}
  \caption{Predicted solubility of QKAN heads across Fourier orders. Each group, including the MLP baseline, contains 100 generated sequences scored with the same external predictor as Figure~\ref{fig:mode-solubility}. Both figures use identical MLP sequences and scores. Violins show Gaussian kernel density estimates with Scott's bandwidth, restricted to the observed score ranges and scaled to equal maximum widths. Dots show individual sequence scores with horizontal jitter. Boxes span the interquartile range (IQR), central bars mark medians, and whiskers extend to the most extreme observations within 1.5 IQR of the box. Diamonds mark means. The dashed horizontal line denotes the MLP mean (0.6088).}
  \label{fig:qkan-order-solubility}
\end{figure}

We next examine how Fourier order relates to predicted solubility across seven simulated QKAN configurations with $K=1,2,3,4,5,6,$ and $8$. The comparison uses the same MLP reference and the same external evaluator, as shown in \cref{tab:qkan-generation-solubility,fig:qkan-order-solubility}.

\begin{table}[!htbp]
  \centering
  \papertablestyle
  \begin{threeparttable}
    \caption{Predicted solubility scores and parameter counts for simulated QKAN heads at different Fourier orders, with the shared MLP reference.}
    \label{tab:qkan-generation-solubility}
    \begin{tabular*}{\linewidth}{@{\extracolsep{\fill}}lcrrrrr@{}}
      \toprule
      & & \multicolumn{3}{c}{\thead{Predicted score}}
      & \multicolumn{2}{c}{\thead{Trainable parameters}} \\
      \cmidrule(lr){3-5}\cmidrule(lr){6-7}
      \thead{Head} & \thead{Fourier order} $\bm{K}$ & \thead{Mean} & \thead{Min.} & \thead{Max.} & \thead{Total} & \thead{QKAN module} \\
      \midrule
      MLP  & --- & 0.6088 & 0.1593 & 0.9197 & 18,649 & --- \\
      \midrule
      \multirow{7}{*}{QKAN} & 1 & 0.4623 & 0.1663 & 0.7284 & 19,145 & 564 \\
           & 2 & 0.4449 & 0.1593 & 0.7084 & 19,225 & 644 \\
           & 3 & 0.6540 & 0.1302 & \underline{0.9285} & 19,305 & 724 \\
           & 4 & 0.6899 & 0.2195 & 0.8887 & 19,385 & 804 \\
           & 5 & 0.6057 & 0.2100 & 0.9202 & 19,465 & 884 \\
           & 6 & \textbf{0.7471} & \textbf{0.5180} & 0.9282 & 19,545 & 964 \\
           & 8 & \underline{0.7143} & \underline{0.2673} & \textbf{0.9377} & 19,705 & 1,124 \\
      \bottomrule
    \end{tabular*}
    \begin{tablenotes}[flushleft]\papertablenotes
      \item[] Bold and underlined values denote the highest and second highest QKAN entries in each score column. Minima and maxima are the observed scores across 100 generated sequences per configuration. Total counts refer to the auxiliary prediction head and include parameters of the QKAN module, derived from the architecture with fixed layer widths in \cref{app:photonic-kan-circuit}.
    \end{tablenotes}
  \end{threeparttable}
\end{table}

The mean predicted solubility varies substantially across Fourier orders. At $K=1$ and $K=2$, the mean scores are 0.4623 and 0.4449, both below the MLP reference of 0.6088. The means increase to 0.6540 and 0.6899 at $K=3$ and $K=4$, respectively, before decreasing to 0.6057 at $K=5$. The best mean score of 0.7471 is achieved at $K=6$, representing a 22.7\% increase relative to the MLP reference.

The QKAN configuration at $K=6$ contains 19,545 trainable head parameters, compared with 18,649 for the MLP, a difference of approximately 4.8\%. For the fixed layer widths, the QKAN module contains $80K+484$ trainable parameters and the complete auxiliary head contains $80K+19{,}065$, including biases (\cref{app:photonic-kan-circuit}). This comparison places the strongest QKAN result near the MLP reference in terms of auxiliary head parameter count.

The distributions in \cref{fig:qkan-order-solubility} show that $K=6$ also has the highest observed minimum score among the QKAN configurations, at 0.5180, compared with 0.1593 for the MLP reference. Its maximum observed score is 0.9282, whereas the highest maximum across the QKAN configurations, 0.9377, occurs at $K=8$. Increasing the order from $K=6$ to $K=8$ raises the total head parameter count from 19,545 to 19,705 but reduces the mean score from 0.7471 to 0.7143. The variation across orders shows that the observed scaling trend is not monotonic.

Taken together, the QNN and QKAN comparisons show that the type and configuration of an auxiliary prediction head are associated with meaningful differences in the predicted solubility of constrained MAO redesigns. Under the fixed generation conditions examined here, selected simulated quantum auxiliary heads achieve higher mean scores than the MLP reference. 

\FloatBarrier
\subsection{Quantum Auxiliary Supervision with Photonic Hardware Responses}
\label{sec:hardware-training}

We further investigate whether photonic hardware responses can support gradient based optimization of quantum auxiliary supervision. These experiments are separate from the numerical comparisons in \cref{sec:trained-model-results}. The QNN experiment uses direct measurements from a photonic quantum computer, whereas the QKAN experiment uses a differentiable response map calibrated from previously measured device probabilities.

\begin{figure}[!htbp]
  \centering
  \includegraphics[width=\linewidth]{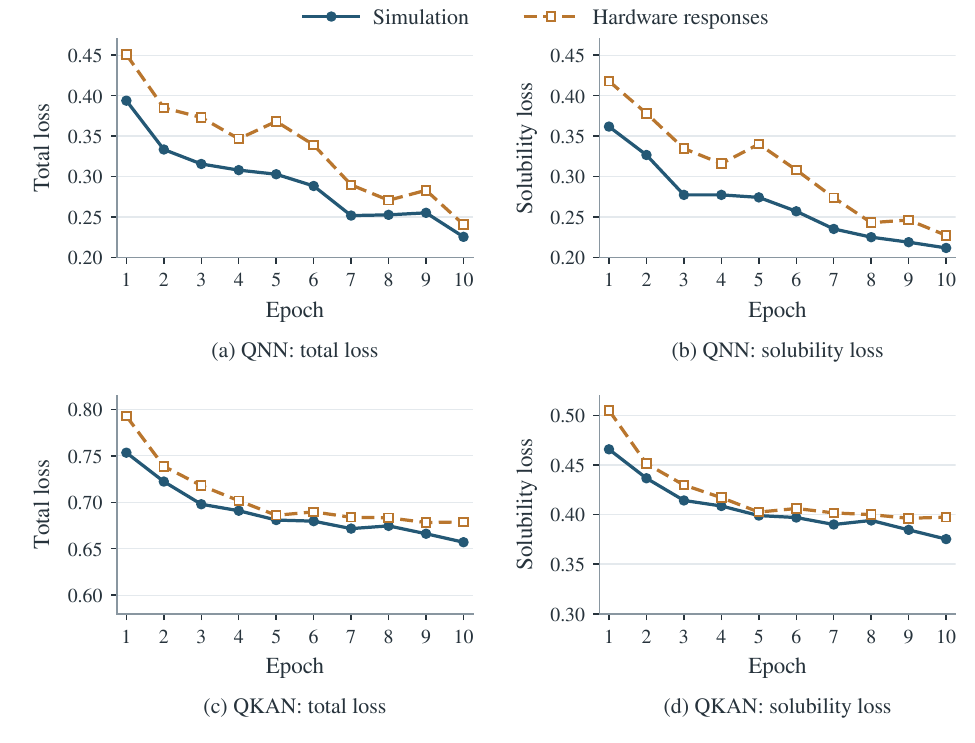}
  \caption{Training using gradients from photonic hardware responses over the first 10 epochs. Panels (a) and (b) compare simulation with direct quantum device measurements for the QNN experiment. Panels (c) and (d) compare simulation with a differentiable response map calibrated from previously measured device probabilities for QKAN.}
  \label{fig:hardware-training}
\end{figure}

For the QNN experiment, we used a Clements photonic circuit to implement unitary transformations on TuringQ's Gen3 photonic quantum computer and a subset of the training data. Expectation values of photonic circuit were obtained directly from the device, and the parameter shift rule provided derivatives with respect to both trainable circuit parameters and encoded input angles. These derivatives were propagated to the classical optimizer during joint training. Over the first 10 training epochs, the total loss decreased from 0.4508 to 0.2407, while the solubility loss decreased from 0.4179 to 0.2272 (\cref{fig:hardware-training}, panels a and b). Both loss trajectories decreased overall, with some fluctuations between epochs.

For QKAN, we examined a separate training run using a differentiable response map constructed from previously measured photon detection probabilities. The calibration data remained fixed during training. Over the first 10 epochs, the total loss decreased from 0.7927 to 0.6788, and the solubility loss decreased from 0.5050 to 0.3973 (\cref{fig:hardware-training}, panels c and d). As in the QNN experiment, both losses decreased overall. For display, the values in panels (c) and (d) are multiplied by 100 and 10, respectively.

Together, these experiments provide preliminary evidence that quantum auxiliary supervision can be optimized using either direct photonic measurements or a response model calibrated from hardware data. The QNN experiment uses device measurements during training, whereas the QKAN experiment does not require live device measurements at each optimization step. Neither experiment should be interpreted as a hardware implementation of the full numerical solubility comparisons reported in \cref{sec:trained-model-results}.

\subsection{Sequence, Structure, and Experimental Functional Validation}
\label{sec:independent-validation}

Separate sequence design tasks generated 1,000 candidates for each of five protein systems: plant defensin (PDB: 1BK8, 50 residues), xylanase (PDB: 2DCZ, 185 residues), short chain dehydrogenase (PDB: 4BMV, 262 residues), transaminase (PDB: 4E3Q, 473 residues), and UDP glycosyltransferase UGT76G1 (UniProt: Q6VAB4, 458 residues). The lengths given here refer to the deposited PDB polymer sequences and, for UGT76G1, the canonical UniProt sequence. Candidates were ranked using sequence identity to the reference, secondary structure similarity, predicted solubility, and conservation of key motifs in multiple sequence alignments (MSAs).

For plant defensin, the 50 selected candidates had a mean predicted secondary structure similarity of 88.24\%, compared with 70.05\% across the full library of 1,000 sequences. Their sequence identities to 1BK8 ranged from 42.00\% to 54.00\%, and all 50 retained the eight cysteine positions aligned to the reference (Table~\ref{tab:system-statistics}).

\begin{table}[!htbp]
  \centering
  \papertablestyle
  \caption{Statistics for generated libraries and selected candidates. For defensin, xylanase, and SDR, sequence identity summarizes the 50 selected candidates, whereas secondary structure similarity summarizes the full libraries of 1,000 sequences.}
  \label{tab:system-statistics}
  \begin{tabularx}{\linewidth}{@{}P{0.12\linewidth}YYY@{}}
    \toprule
    \thead{System} & \thead{Sequence identity (\%)} & \thead{Secondary structure similarity (\%)} & \thead{Predicted solubility} \\
    \midrule
    Defensin & 42.00--54.00 (mean 46.32) & 16.00-92.00 (mean 70.05) & 0.194--0.859 (mean 0.634) \\
    Xylanase & 49.45--57.71 (mean 53.55) & 45.41-90.27 (mean 68.97) & 0.210--0.791 (mean 0.466) \\
    SDR & 35.23--42.80 (mean 38.43) & 41.96-83.53 (mean 69.83) & 0.159--0.620 (mean 0.350) \\
    ATA & 49.55--53.65 (mean 51.40) & 74.33-82.74 (mean 78.43) & 0.207--0.826 (mean 0.469) \\
    UGT & 41.47-45.66 (mean 43.56) & 70.96-78.16 (mean 74.04) & 0.337--0.729 (mean 0.586) \\
    \bottomrule
  \end{tabularx}
\end{table}

The 50 selected xylanase candidates had a mean predicted secondary structure similarity of 83.33\% and a mean BLAST identity of 53.55\% to 2DCZ. All retained the GGNY, GKGW, YGWT, NAP, WSVR, and ATEG motifs.

\begin{figure}[!htbp]
  \centering
  \includegraphics[width=\linewidth]{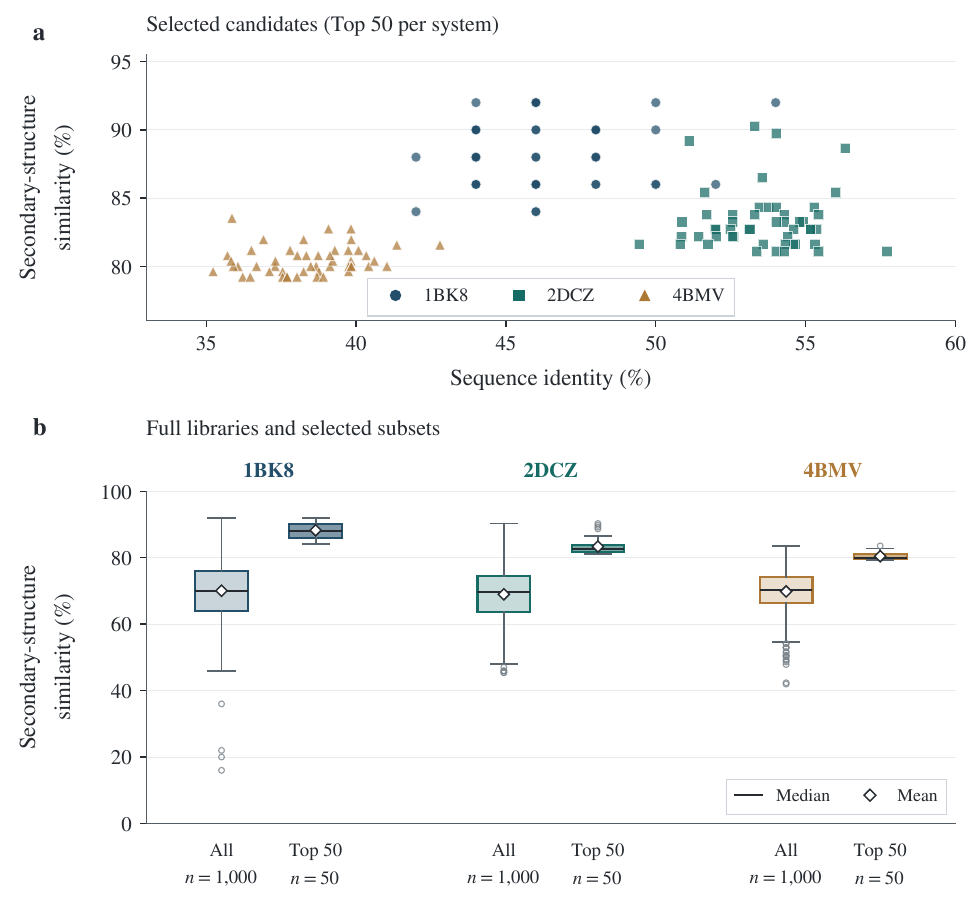}
  \caption{Sequence and secondary structure evaluation for defensin (1BK8), xylanase (2DCZ), and SDR (4BMV).
  (a) Sequence identity versus predicted secondary structure similarity for the 50 selected candidates per system. Coincident points overlap.
  (b) Secondary structure similarity distributions for each full library ($n=1{,}000$) and its selected Top 50 subset ($n=50$).
  Boxes span the interquartile range (IQR), horizontal lines denote medians, and diamonds denote means.
  Whiskers extend to the most extreme observations within 1.5 IQR of the box. Observations beyond the whiskers are shown individually.}
  \label{fig:identity-structure}
\end{figure}

For SDR, the 50 selected candidates had a mean predicted secondary structure similarity of 80.44\%, compared with 69.83\% across the full library, and a mean BLAST identity of 38.43\% to 4BMV. All retained the TGAS/GIGA segment associated with the Rossmann fold and the NNAG and QAVL motifs. Figure~\ref{fig:identity-structure} compares the sequence identities and secondary structure similarity distributions for these three systems.

\begin{table}[!htbp]
  \centering
  \papertablestyle
  \caption{Experimental activities of selected protein candidates.}
  \label{tab:validation}
  \begin{tabular}{@{}P{0.15\linewidth}P{0.15\linewidth}P{0.25\linewidth}P{\dimexpr0.45\linewidth-6\tabcolsep\relax}@{}}
    \toprule
    \thead{System} & \thead{Validated candidate} & \thead{Assay} & \thead{Result} \\
    \midrule
  Plant defensin & Defensin A & Aspergillus niger inhibition zone assay & Approximately 10 mm at 0, 12, and 24 h \\
  Plant defensin & Defensin B & Aspergillus niger inhibition zone assay & Approximately 12--13 mm at 0, 12, and 24 h (strongest among the three) \\
  Plant defensin & Defensin C & Aspergillus niger inhibition zone assay & Approximately 10 mm at 0 and 12 h and approximately 8 mm at 24 h \\
  Xylanase & Xyn1 & DNS specific activity assay & 1.93 $\pm$ 0.11 U/mg, with 100\% residual activity after 55~$^{\circ}$C for 2 h \\
  Xylanase & Xyn2 & DNS specific activity assay & 2.21 $\pm$ 0.09 U/mg, with 100\% residual activity after 55~$^{\circ}$C for 2 h \\
  Xylanase & Xyn3 & DNS specific activity assay & 2.17 $\pm$ 0.118 U/mg, with 100\% residual activity after 55~$^{\circ}$C for 2 h \\
  Xylanase & Xyn4 & DNS specific activity assay & 1.92 $\pm$ 0.093 U/mg, with 100\% residual activity after 55~$^{\circ}$C for 2 h \\
  Short chain dehydrogenase & SDR1 & specific activity assay with NADP\textsuperscript{+} & 55.25 $\pm$ 3.53 U/mg \\
  Short chain dehydrogenase & SDR2 & specific activity assay with NADP\textsuperscript{+} & 53.10 $\pm$ 1.41 U/mg \\
    \bottomrule
  \end{tabular}
\end{table}

Functional assays showed that the tested defensins inhibited \textit{Aspergillus niger}, while the xylanase and SDR candidates exhibited enzymatic activity (Table~\ref{tab:validation}). Defensin B produced the largest inhibition zones among the three tested candidates. Xyn1--Xyn4 retained 100\% residual activity after incubation at 55~$^{\circ}$C for 2 h.

\section{Conclusion}

DiffSequence-Q combines multinomial diffusion conditioned on protein secondary structure and solubility with jointly trained quantum property predictors. The QNN and QKAN predictors introduce photonic interference features and structured Fourier responses, respectively, while joint supervision updates the shared denoising representation. On the CATH benchmark, DiffSequence-Q achieved the highest secondary structure recovery among the four evaluated models for complete sequence generation (Table~\ref{tab:protein-generation}). In the solubility evaluation, the best QNN and QKAN configurations achieved higher mean predicted solubility scores than the classical MLP reference (Tables~\ref{tab:solubility} and~\ref{tab:qkan-generation-solubility}).

We also evaluated training with photonic quantum hardware responses. The QNN experiment used direct circuit measurements on TuringQ's Gen3 photonic quantum computer, whereas the QKAN experiment used a differentiable response map calibrated from previously measured photonic quantum probabilities. Both experiments reduced the total and solubility training losses over the first 10 epochs, supporting hybrid optimization with responses derived from hardware.

Functional and biochemical assays demonstrated biological activity in selected defensins, xylanases, and short-chain dehydrogenases generated by the model, extending the evaluation beyond computational predictions to antimicrobial and enzymatic activity. Together, these results support DiffSequence-Q as a framework for integrating structural constraints and solubility targets in functional protein design. Future studies will extend experimental validation to quantitative yields of soluble protein expression, fold stability, and structural characterization across recalcitrant protein targets.

\section{Acknowledgements}
\label{sec:ack}

This research is supported by the National Key R\&D Program of
China (Grant No.~2024YFB4504005 and 2024YFA1409300);
National Natural Science Foundation of China (NSFC)
(Grant Nos.~62235012, 12304342, 12574549, 12574542, and 125B1033);
Innovation Program for Quantum Science and Technology
(Grant Nos.~2021ZD0301500 and 2021ZD0300700);
Science and Technology Commission of Shanghai Municipality
(STCSM) (Grant Nos.~2019SHZDZX01, 24ZR1438700, 24ZR1430700,
and 24LZ1401500);
Shanghai Municipal Commission of Economy and Informatization (no. 2025-GZL-RGZN-02047);
and Frontier Technologies R\&D Program of Jiangsu
(Grant No.~SBF20250000094).

\clearpage

\FloatBarrier
\appendix

\section{Photonic Quantum Circuits}
\label{app:circuit}

\subsection{QNN head}
\label{app:four-mode-unitary}

The QNN realization of $\mathcal H_\omega$ in \cref{sec:solubility-heads}
maps the pooled representation $\mathbf u^{(b)}$ to $m-1$ relative phase angles
through an input encoder. Its quantum module is a circuit with $m$ optical
modes and one photon, whose output features $\mathbf z^{(b)}$ are mode
occupation probabilities. The scalar prediction is
$\widehat y^{(b)}=\mathbf w_o^\top\mathbf z^{(b)}+b_o$, where $\mathbf w_o$ and
$b_o$ are the affine readout parameters. This section specifies the quantum
module using four optical modes.

\begin{figure}[htbp]
  \centering
  \includegraphics[width=\textwidth]{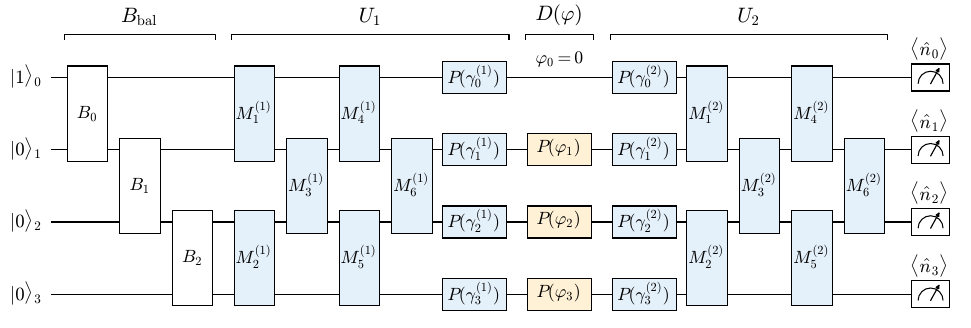}
  \caption{Photonic circuit using one photon in four modes, with fixed state preparation,
  two trainable Clements meshes, and a single layer of phase encoding.
  Blue gates are trainable. Yellow gates encode data.
  $B_j$, $M_k^{(s)}$, and $P$ denote fixed beam splitters,
  Mach--Zehnder interferometers, and phase shifters, respectively.}
  \label{fig:four_mode_circuit}
\end{figure}

Starting from $|\psi_{\mathrm{in}}\rangle=|1,0,0,0\rangle$, the circuit
preserves the subspace with one photon. Circuit matrices use the basis
$\{|1_j\rangle=\hat a_j^\dagger|\mathrm{vac}\rangle\}_{j=0}^{3}$.
The data angles $\mathbf d=(d_1,d_2,d_3)$ appear as $\varphi_j=d_j$ in
\cref{fig:four_mode_circuit}, with reference phase $\varphi_0=0$.
The trainable angles are
\begin{equation}
 \begin{aligned}
  \phi&=(\phi^{(1)},\phi^{(2)})\in\R^{32},\\
  \phi^{(s)}&=\bigl((\theta_k^{(s)},\eta_k^{(s)})_{k=1}^{6},
                   (\gamma_j^{(s)})_{j=0}^{3}\bigr)\in\R^{16},
                   \qquad s\in\{1,2\}.
 \end{aligned}
\end{equation}
Let $U_1(\phi^{(1)})$ and $U_2(\phi^{(2)})$ denote the first and second
trainable interferometer matrices, $B_{\mathrm{bal}}$ the fixed matrix for state
preparation, and $D(\mathbf d)$ the matrix for data encoding.
The total circuit unitary $U_\phi(\mathbf d)\in\mathbb C^{4\times4}$
acts on amplitudes in the subspace with one photon,
with operators applied from right to left:
\begin{equation}
 U_\phi(\mathbf d)=
 U_2(\phi^{(2)})\,D(\mathbf d)\,U_1(\phi^{(1)})\,B_{\mathrm{bal}}.
 \label{eq:circuit-unitary}
\end{equation}

The fixed preparation is $B_{\mathrm{bal}}=B_2B_1B_0$, where
$B_j=\operatorname{BS}_{j,j+1}(\vartheta_j,0)$ and
$\vartheta_j=\arccos(1/\sqrt{4-j})$ for $j=0,1,2$.
These angles give an occupation probability of $1/4$ in each output
mode. The signs of the amplitudes also depend on the splitter convention.
In the ordered basis $(|1_j\rangle,|1_{j+1}\rangle)$, consider the
real rotation
\[
 B_j\big|_{j,j+1}=
 \begin{pmatrix}
 \cos\vartheta_j & -\sin\vartheta_j\\
 \sin\vartheta_j &  \cos\vartheta_j
 \end{pmatrix}.
\]
With this convention,
$B_{\mathrm{bal}}|\psi_{\mathrm{in}}\rangle
=\tfrac12\sum_{j=0}^{3}|1_j\rangle$.
With $P_j(\chi)=\exp(i\chi\hat n_j)$ and
$\hat n_j=\hat a_j^\dagger\hat a_j$, data are encoded once through
\begin{equation}
 D(\mathbf d)=\prod_{j=1}^{3}P_j(d_j)
 =\diag\bigl(1,e^{id_1},e^{id_2},e^{id_3}\bigr).
 \label{eq:four-mode-encoding}
\end{equation}

Each Clements mesh contains six MZIs with two parameters each and four trainable phase
shifters. MZI layers 1 and 3 couple $(0,1)$ and $(2,3)$ in parallel.
Layers 2 and 4 couple $(1,2)$.
Defining $\mathcal M_s=M_6^{(s)}\cdots M_1^{(s)}$ and
$S_s=\prod_{j=0}^{3}P_j(\gamma_j^{(s)})$, their ordering is
\begin{equation}
 \begin{aligned}
  U_1(\phi^{(1)})&=S_1\mathcal M_1,\\
  U_2(\phi^{(2)})&=\mathcal M_2S_2.
 \end{aligned}
 \label{eq:four-mode-meshes}
\end{equation}

The local MZI matrix in the first mesh is
\begin{equation}
 \mathsf M(\theta,\eta)=ie^{i\theta/2}
 \begin{pmatrix}
  e^{i\eta}\sin(\theta/2) & \cos(\theta/2)\\
  e^{i\eta}\cos(\theta/2) & -\sin(\theta/2)
 \end{pmatrix}.
 \label{eq:photonic-mzi}
\end{equation}
Each $M_k^{(1)}$ embeds this $2\times2$ block on its indicated mode pair
and acts as the identity on the other modes. Each $M_k^{(2)}$ instead
embeds $\mathsf M(\theta_k^{(2)},\eta_k^{(2)})^{\mathsf T}$, using
independent parameters. The superscript $\mathsf T$ denotes ordinary transpose.

For $|\psi_{\mathrm{out}}\rangle=U_\phi(\mathbf d)|\psi_{\mathrm{in}}\rangle$,
the four readout features are
\begin{equation}
 e_j(\mathbf d,\phi)
 =\langle\psi_{\mathrm{out}}|\hat n_j|\psi_{\mathrm{out}}\rangle
 =\bigl|\langle1_j|\psi_{\mathrm{out}}\rangle\bigr|^2,
 \qquad j=0,1,2,3.
 \label{eq:four-mode-readout}
\end{equation}
These expected photon numbers are mode occupation probabilities,
with $0\le e_j\le1$ and $\sum_{j=0}^{3}e_j=1$.

\subsection{QKAN head}
\label{app:photonic-kan-circuit}

The QKAN realization of $\mathcal H_\omega$ projects the pooled
representation through an input encoder with dimensions $128\to128\to16$.
Three QKAN layers with channel dimensions $16\to16\to8\to4$ transform the
resulting features, followed by a $4\to1$ affine readout. Each QKAN layer
combines univariate Fourier responses with learned linear channel mixing.
For input channel $i$ within a layer, the response is parameterized as
\begin{equation}
 \begin{aligned}
  f_i(x_i)&=c_i+\sum_{k=1}^{K}
  \bigl[a_{ik}\cos(k\varphi_i)+b_{ik}\sin(k\varphi_i)\bigr],\\
  \varphi_i&=\frac{2\pi x_i}{P},\qquad P=2\pi.
 \end{aligned}
 \label{eq:kan-fourier-response}
\end{equation}
Here $K$ is the Fourier order, and $c_i$, $a_{ik}$, and $b_{ik}$ are
trainable coefficients.

\begin{figure}[!htbp]
  \centering
  \includegraphics[width=0.94\linewidth]{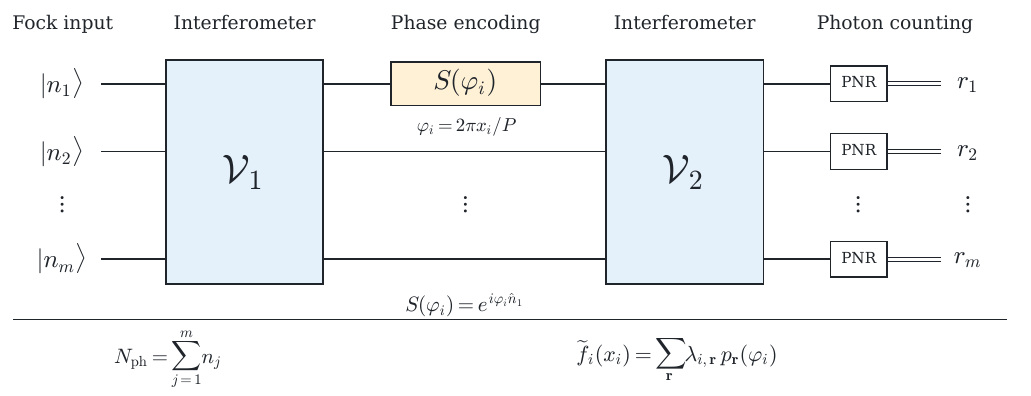}
  \caption{Schematic of a photonic quantum Fourier circuit, following
  Gan et al.~\citep{Gan2022FockExpressivity}.
  A single phase shifter encodes the input between two interferometers.
  Wires denote optical modes. PNR denotes detection that resolves photon numbers.}
  \label{fig:qkan_circuit}
\end{figure}

Figure~\ref{fig:qkan_circuit} shows an architecture based on linear optics for
constructing univariate Fourier responses, following
Gan et al.~\citep{Gan2022FockExpressivity}. We describe its ideal state
evolution, probabilities of photon counts, and weighted readout.

For input channel $i$, the encoded phase $\varphi_i$ is defined in
\cref{eq:kan-fourier-response}. A Fock input over $m$ modes
$|\mathbf n_{\mathrm{in}}\rangle=|n_1,\ldots,n_m\rangle$ with
$N_{\mathrm{ph}}=\sum_{j=1}^{m}n_j$ photons evolves under the ideal,
lossless unitary
\begin{equation}
 \begin{aligned}
  \mathcal U(\varphi_i)&=\mathcal V_2 S(\varphi_i)\mathcal V_1,\\
  S(\varphi_i)&=\exp(i\varphi_i\hat n_1).
 \end{aligned}
 \label{eq:kan-photonic-unitary}
\end{equation}
Here $\mathcal V_1$ and $\mathcal V_2$ represent interferometers on Fock space
that preserve photon number, and $\hat n_1$ counts photons in the first mode.
Detection that resolves photon numbers yields
\begin{equation}
 \begin{aligned}
  p_{\mathbf r}(\varphi_i)
   &=\bigl|\langle\mathbf r|\mathcal U(\varphi_i)
                  |\mathbf n_{\mathrm{in}}\rangle\bigr|^2,\\
  \widetilde f_i(x_i)
   &=\sum_{\mathbf r}\lambda_{i,\mathbf r}\,p_{\mathbf r}(\varphi_i).
 \end{aligned}
 \label{eq:kan-probability-readout}
\end{equation}
The sum runs over output occupations with $\sum_j r_j=N_{\mathrm{ph}}$,
and $\lambda_{i,\mathbf r}\in\R$ are observable weights.
Since the encoded mode has occupation numbers $0,\ldots,N_{\mathrm{ph}}$,
the phase differences contributing to the probabilities have integer
frequencies between $-N_{\mathrm{ph}}$ and $N_{\mathrm{ph}}$. The readout is real and therefore admits the finite Fourier expansion
\begin{equation}
 \widetilde f_i(x_i)=\widetilde c_i+
 \sum_{k=1}^{N_{\mathrm{ph}}}
 \bigl[\widetilde a_{ik}\cos(k\varphi_i)
       +\widetilde b_{ik}\sin(k\varphi_i)\bigr].
 \label{eq:kan-circuit-fourier-response}
\end{equation}
The input state, interferometer configuration, and observable weights
jointly determine these coefficients, some of which may vanish.
A nonzero $K$th harmonic requires $N_{\mathrm{ph}}\ge K$.
Matching a specified response in \cref{eq:kan-fourier-response} additionally
requires that its Fourier coefficients be realizable by the chosen circuit
and readout. Frequency coverage and coefficient realizability are distinct
conditions\citep{Gan2022FockExpressivity}.

\FloatBarrier
\subsection{Gradient propagation through the quantum modules}
\label{app:property-gradient}

The quantum modules of the QNN and QKAN heads share the
following formulation based on the chain rule. For a module with data angles $\boldsymbol\alpha$
and trainable parameters $\boldsymbol\eta$, let
$\mathbf z=\mathbf q(\boldsymbol\alpha,\boldsymbol\eta)$ denote its readout.
If $\boldsymbol\alpha=A(\mathbf v)$ encodes the preceding representation
$\mathbf v$, and $\overline{\mathbf z}=\partial\ell_y/\partial\mathbf z$
is the upstream gradient, then
\begin{align}
 \nabla_{\mathbf v}\ell_y
 &=J_{\boldsymbol\alpha,\mathbf v}^{\top}
   J_{\mathbf z,\boldsymbol\alpha}^{\top}\overline{\mathbf z},
 \label{eq:full-property-vjp}\\
 \nabla_{\boldsymbol\eta}\ell_y
 &=J_{\mathbf z,\boldsymbol\eta}^{\top}\overline{\mathbf z}.
 \label{eq:circuit-parameter-gradient}
\end{align}
Here $J_{\mathbf a,\mathbf b}=\partial\mathbf a/\partial\mathbf b$, and
the module parameters $\boldsymbol\eta$ are included in the head parameters
$\omega$.
These products are applied successively when several modules are composed.
For ideal circuit readouts, phase derivatives can be evaluated by the shift rules in
\cref{app:photonic-parameter-shift}. Derivatives of the observable weights
follow directly from their linear dependence in the readout.

Consider one labeled protein at its sampled step $t$. In the following
derivation we suppress the indices $b$ and $t$. Write $\mathbf h_\ell$ for
$\mathbf h_{\Theta,\ell}^{(b)}$ in \cref{eq:shared-representation},
$\mathbf u$ for its pooled sequence representation, and
$\ell_y=(\widehat y-y)^2$ for its squared error before weighting by noise level
in \cref{eq:batch-solubility-mse}.
The readout derivative is $\partial\ell_y/\partial\widehat y
=2(\widehat y-y)$. With the validity mask fixed, let
$N_{\mathrm{v}}=\max(1,\sum_\ell m_\ell)$ and
$\mathbf u=N_{\mathrm{v}}^{-1}\sum_\ell m_\ell\mathbf h_\ell\in\R^{128}$.
The derivative through masked mean pooling is
\begin{equation}
 \frac{\partial\ell_y}{\partial\mathbf h_\ell}
 =\frac{m_\ell}{N_{\mathrm{v}}}\nabla_{\mathbf u}\ell_y .
 \label{eq:pool-gradient}
\end{equation}
For the shared encoder parameters $\Theta_{\mathrm{enc}}$, this gives
\begin{equation}
 \nabla_{\Theta_{\mathrm{enc}}}\ell_y
 =\sum_\ell
   \left(\frac{\partial\mathbf h_\ell}{\partial\Theta_{\mathrm{enc}}}\right)^\top
   \frac{m_\ell}{N_{\mathrm{v}}}\nabla_{\mathbf u}\ell_y .
 \label{eq:photonic-encoder-gradient}
\end{equation}
In the batch objective, example $b$ contributes
$(\lambda_{\mathrm{sol}}w^{(b)}/B)\nabla_{\Theta_{\mathrm{enc}}}\ell_y^{(b)}$
to the gradient in \cref{eq:generator-gradient}.
Detaching the shared representation blocks this contribution to the
encoder while preserving the gradients of the auxiliary head.

\subsection{Differentiation using parameter shifts}
\label{app:photonic-parameter-shift}

For the QNN module in \cref{app:four-mode-unitary}, the operator for the subspace with one photon
$\hat n_j=|1_j\rangle\langle1_j|$ has
eigenvalues 0 and 1. The phase shifter $P_j(\xi)=\exp(i\xi\hat n_j)$
therefore satisfies the parameter shift rule for a generator with two
distinct eigenvalues\citep{Schuld2019ParameterShift}.
For the MZIs, define the local matrices
\begin{equation}
 C=\frac{1}{\sqrt2}\begin{pmatrix}1&i\\i&1\end{pmatrix},
 \qquad
 Q(\xi)=\begin{pmatrix}e^{i\xi}&0\\0&1\end{pmatrix}.
\end{equation}
\cref{eq:photonic-mzi} factorizes as
\begin{equation}
 \begin{aligned}
  \mathsf M(\theta,\eta)&=C Q(\theta)C Q(\eta),\\
  \mathsf M(\theta,\eta)^{\mathsf T}&=Q(\eta)C Q(\theta)C.
 \end{aligned}
 \label{eq:mzi-phase-factorization}
\end{equation}
Each MZI angle therefore controls a single phase shifter. For any data
or trainable circuit angle $\xi$, holding all other angles fixed,

\begin{equation}
 \frac{\partial e_j}{\partial\xi}
 =\frac{e_j(\xi+\pi/2)-e_j(\xi-\pi/2)}{2},
 \qquad j=0,1,2,3.
 \label{eq:photonic-parameter-shift}
\end{equation}

The ideal photonic Fourier response in
\cref{eq:kan-circuit-fourier-response} can contain several nonzero frequencies.
For a response $F(\varphi)$ supported on integer frequencies
$-R,\ldots,R$, a generalized parameter shift rule is
\citep{Wierichs2022GeneralShift}
\begin{equation}
 \begin{aligned}
  \frac{\partial F}{\partial\varphi}
   &=\sum_{r=1}^{R}a_r\bigl[F(\varphi+s_r)-F(\varphi-s_r)\bigr],\\
  s_r&=\frac{(2r-1)\pi}{2R},\qquad
  a_r=\frac{(-1)^{r-1}}{4R\sin^2(s_r/2)}.
 \end{aligned}
 \label{eq:fourier-parameter-shift}
\end{equation}
For encoding through a single phase in \cref{eq:kan-photonic-unitary},
$R=N_{\mathrm{ph}}$ bounds the frequencies. The ideal response of order $K$ in \cref{eq:kan-fourier-response} uses $R=K$, requiring $K$
pairs of shifted evaluations. Setting $R=1$ recovers
\cref{eq:photonic-parameter-shift}. Input derivatives include the phase
scaling $\partial\varphi_i/\partial x_i=2\pi/P$.

Parameter shifts are applied to circuit phases while all other parameters remain fixed. In vector--Jacobian products, the upstream gradient is evaluated at the unshifted parameters. These identities apply to ideal circuit expectation values and exact Fourier responses with the stated finite frequency support. Estimates obtained from finitely many measurements are subject to sampling uncertainty. Classical simulation of the ideal circuit also permits automatic differentiation in reverse mode.

\FloatBarrier
\section{Solubility Prediction Model}
\label{app:solubility-evaluator}

\subsection{Model architecture}

We trained an external solubility prediction model on the public DeepSol
dataset~\citep{Khurana2018DeepSol} using frozen embeddings from
ESM++ Large~\cite{Hallee2024ESMplusplus}. A Transformer classifier with
attention pooling and a feedforward readout produces a continuous score
in $[0,1]$ through a sigmoid output. The same trained model was used to
score all generated sequences.

\subsection{Evaluation of continuous scores}

The model was evaluated on a fixed subset of 6,694 sequences from the
DeepSol \textit{E.~coli} dataset and achieved a binary classification
accuracy of 0.81 at a threshold of 0.5.

\FloatBarrier

\renewcommand{\refname}{References}
\bibliographystyle{unsrtnat}
\bibliography{references}

\end{document}